\RequirePackage{fix-cm}
\pdfoutput=1
\documentclass[twocolumn]{svjour3}
\usepackage{amsmath}
\usepackage{graphicx,psfrag,epsf}
\usepackage{enumerate}
\usepackage{natbib}
\usepackage{url} 

\usepackage{amsmath}

\usepackage{amssymb}   
\usepackage{float}         
\usepackage{subcaption}
\usepackage{rotating}
\usepackage{algorithm2e}
\usepackage{wrapfig}
\usepackage{bbold}
\usepackage{multirow}
\usepackage{amsfonts} %
\usepackage{adjustbox}
\usepackage{bm}
\usepackage{verbatim}
\usepackage{mathrsfs}

\newcommand{\bs}[1]{\boldsymbol{#1}}
\DeclareMathOperator*{\argmin}{arg\,min}

\SetAlCapFnt{\normalfont}

\journalname{}

\begin{document}

  \title{\bf Semi-automated simultaneous predictor selection for Regression-SARIMA models}

\author{Aaron Lowther \and Paul Fearnhead \and Matthew A. Nunes \and Kjeld Jensen}

\institute{
 Aaron Lowther \at
  STOR-i Centre for Doctoral Training, Lancaster University, LA1 4YF, United Kingdom
           \and
           Paul Fearnhead \at Department of Mathematics and Statistics, Lancaster University, LA1 4YF, United Kingdom
           \and
           Matthew A. Nunes \at School of Mathematics, University of Bath, BA2 7AY, United Kingdom \\ \email{m.a.nunes@bath.ac.uk}
           \and
           Kjeld Jensen \at BT Applied Research, BT Plc, EC1A 7AJ, United Kingdom
}

\authorrunning{A. Lowther et al.} 

\date{}

\maketitle

\begin{abstract}
Deciding which predictors to use plays an integral role in deriving statistical models in a wide range of applications. Motivated by the challenges of predicting events across a telecommunications network, we propose a semi-automated, joint model-fitting and predictor
selection procedure for linear regression models. Our approach can model and account for serial correlation in the regression residuals, produces sparse and interpretable models and can be used
to jointly select models for a group of related responses. This is achieved through fitting linear models under constraints on the number of non-zero coefficients using a generalisation of a recently developed 
Mixed Integer Quadratic Optimisation approach. The resultant models from our approach achieve better predictive performance on the motivating telecommunications data than methods currently used by industry.
\end{abstract}

\keywords{best subset selection; linear regression; mixed integer quadratic optimisation; multivariate response model.}

\section{Introduction} \label{sec:introduction}

The use of statistical models to drive business efficiency is becoming increasingly widespread \citep{BidDataDecisions}. Consequently, organisations are recording more and more data for subsequent analysis (see \cite{BigData} or \cite{MachineLearningTrends} for a review of current modelling challenges in this area). As a result, traditional (\textit{manual}) approaches for building statistical models are often infeasible for the ever-increasing volumes of data. Automating these approaches is thus necessary, and will allow principled statistical methods to continue being at the forefront of business practice.  


The work in this article is motivated by modelling challenges faced by an industrial collaborator.  The data we consider consists of daily event observations from multiple locations within a telecommunications network. The task of interest is to develop models for how the rates of these events depend on a range of external factors so as to better understand the physical relationship between the network and external influences.  The number of such predictors of events considered for a model in this setting can be in the tens or hundreds, and often it is natural to choose candidates within groups of predictors.  Whilst historically practitioners have fitted such models by hand, this is costly.  The statistical challenge in this context is therefore to fit sparse and interpretable models for the responses, whilst accounting for the serial correlation in the data and ensuring we borrow information across the response variables.  This modelling task needs to be accomplished with minimal human input.  

A body of work in the statistical literature is devoted to predictor selection in univariate response models, see for example, \cite{Hocking1976, Tibshirani1996, ZouHastie2005_ElasticNet, BerstimasKingMazumder2016} and \cite{HastieTibshiraniTibshirani2017} and the references therein. \cite{HastieTibshiraniFriedmanESL} provide an accessible review of many of these methods.  In the multiple response setting, \cite{BreimanFriedman1997} and \cite{SrivastavaSolanky2003} have shown that simultaneous model estimation has advantages over individual modelling procedures. \cite{TurlachVenablesWright2005}, \cite{SimiliaTikka2007} and \cite{SimonFriedmanHastie2013} consider selecting variables for the multi-response models used by \cite{BreimanFriedman1997} and \cite{SrivastavaSolanky2003}. 

In our setting, due to the grouped nature and large number of potential predictors, it is natural to consider combinatorial approaches to predictor selection.  To this end, in this article we propose a multivariate response implementation of the so-called {\em best subset problem} \citep{Miller2002}, and perform predictor selection via a generalisation of the Mixed Integer Quadratic Optimisation (MIQO) model approach of \cite{BerstimasKingMazumder2016}. To the best of our knowledge, addressing the task of simultaneous predictor selection for multiple separate linear regression models via a MIQO formulation has not been considered in the literature.  

Our approach is to fit the same model form for each response variable, but allow for the coefficients associated with a particular predictor to vary across each model. We expand the scope of the original MIQO formulation to automatically fit such a model in the presence of a known serial correlation structure for the time series of responses by considering more general regression seasonal autoregressive integrated moving average (Reg-SARIMA) models, and propose an iterative procedure that alternates between learning the serial correlation structure and fitting the model. We find that a more accurate specification of the model for the regression residuals can lead to a significant reduction in the variance of the predictor selection routine. Using the generalised least squares objective \citep{RaoToutenburg-LinearModels} we can improve inference and predictor selection accuracy.  

To improve model sparsity, our approach can also shrink the coefficients associated with a particular predictor to a common value if desired. The model fitting can be performed under constraints that avoid including highly correlated predictors, which increases the interpretability of the final models. 
Hence with our proposed semi-automated procedure, we reduce the human input by modelling characteristics of the response variables, instead of determining subjective pre-processing steps to remove this variation. The only user input needed is through choosing an appropriate set of initial predictors and potential non-linear transformations of these variables. Here, we estimate the serial correlation by pre-specifying a suitable list of time series models, although iterative approaches adopted by \cite{HyndmanKhandakar_forecast} could be incorporated very easily. Our implementation is computationally feasible for hundreds of predictors and multiple response variables. 

This article is structured as follows. In Section \ref{sec:problem-statement} we review pertinent literature for predictor selection and propose how to use the formulations of \cite{BertsimasKing2016} to develop an automated modelling procedure. In Section \ref{sec:simult-pred-select} we introduce our multi-response MIQO formulation and extensions that can improve the performance of the models. In particular, Section \ref{sec:appl-seri-corr} outlines our two-step procedure which can perform predictor selection whilst accounting for serial correlation in the data.  Section \ref{sec:simulation-study} highlights the advantages of our approach over standard methods in the literature through a simulation study. We apply our approach to a motivating data application in Section \ref{sec:data-study} before concluding the article in Section \ref{sec:concl-furth-work}. 

\section{Problem statement and existing approaches} \label{sec:problem-statement}
In this section we first review the standard linear regression model and existing methods for choosing suitable predictors. We then outline how we propose to automate modelling for one response variable and show how expert opinion can be incorporated into the model.

The linear regression model is able to describe the relationship between a response variable, $Y$ and dependent variables, $X_1, \ldots, X_P$ as follows:
\begin{align}
  \label{eq:linear-regression-model}
  Y = \sum_{p=1}^{P} X_{p} \beta_p + \eta,
\end{align}
where $\eta$ is assumed to be normally distributed, $\eta\sim N(0,\sigma^2_\eta)$. If the set of predictors $\mathcal{X}:=\{X_1, \ldots, X_P\}$ is known, the coefficients $\bs{\beta} = [\beta_1, \dots, \beta_P]$ can be estimated with the standard ordinary least squares (OLS) estimate \begin{align}
  \label{eq:OLS}
  \hat{\bs{\beta}}_{OLS} = \argmin_{\bs{\beta}}
  \left\{ \sum_{t=1}^{T} \left( y_t - \sum_{p=1}^{P}X_{t,p} \beta_{p} \right)^2 \right\}.
\end{align}
When $P$ is large and $\mathcal{X}$ contains redundant predictors, OLS estimates can be unsatisfactory. Prediction accuracy can be improved by shrinking or setting some of the coefficients to zero \citep{HastieTibshiraniFriedmanESL}. Setting coefficients to zero removes the corresponding predictors from \eqref{eq:linear-regression-model}, leading to simpler, more interpretable models. Throughout this article we refer to the number of non-zero coefficients in the model as the {\em model sparsity}, which we denote by $k$.

The regression model above assumes a {\em linear} relationship between predictors and a response variable but this may not be suitable \citep{RawlingsPantulaSastryDickey}. For instance, in our motivating example some telecommunication events are caused by long periods of heavy rainfall, causing underground cables to flood. Exponential smoothing can be applied to daily precipitation measurements to provide a surrogate predictor for ground water levels. This introduces the question of how best to choose the smoothing parameter.  One option is to obtain such surrogate predictors for a grid of smoothing parameters; this both substantially increases the number of potential predictors to choose from, and can lead to highly correlated predictors.  We note here that in other contexts, different transformed variables could be appropriate, for example models which include lagged predictors.

In this article, we focus on subset selection methods that attempt to choose the set of $k$ predictors that give the smallest value of the residual sum of squares \eqref{eq:OLS}. 
A number of classical subset methods are described in detail by \cite{Hocking1976}. The forward-stepwise routine is the current algorithm of choice for selecting predictors by our industrial collaborator. This algorithm is usually initialised with an intercept term (the null model), and iteratively adds the predictor which most improves the least squares objective. This gives a fitted model with $k$ predictors, for some $k\in\{1,\ldots,P\}$. However, for any $k \geq 2$ the model produced by stepwise methods is not guaranteed to be the best model with $k$ predictors in terms of minimising the least squares objective. Despite the resultant sub-optimal models and issues raised by many authors, e.g.\ \cite{Beale1970, Mantel1970, Hocking1976, Berk1978}, fast and easy implementation of these algorithms may explain why they remain popular.

Finding the model with sparsity $k$ which minimises the least squares objective is known as the \textit{best subset} problem \citep{Miller2002}. This optimisation problem is non-convex, and can be computationally challenging to solve when we have many predictors available. However, \cite{BerstimasKingMazumder2016} show that by appropriately formulating the problem and using
recent developments in optimisation algorithms it is possible to perform best subset selection with
hundreds of potential predictors and thousands of observations. \cite{BerstimasKingMazumder2016} also show that best subset selection tends to produce sparser and more interpretable models than more computationally efficient procedures such as the LASSO \citep{Tibshirani1996}.

\subsection{Automated predictor selection procedure} \label{sec:our-prop-autom}
Automated model selection procedures limit an analyst's control over the output. Consequently, we do not seek a fully automated approach, but one that can produce sensible outputs with minimal user input for potentially hundreds of predictors.  We thus propose a semi-automated procedure where an analyst supplies a suitable set of predictors, with which we use best subset selection to automatically choose the best model using this set.  

We formulate the problem of choosing the best model as a Mixed Integer Quadratic Optimisation (MIQO) as suggested by \cite{BerstimasKingMazumder2016}.  The MIQO formulation with sparsity $k$ solves the minimisation 
\begin{subequations}\label{eq:univariate-formulation}
  \begin{align}
    \min_{\bs{\beta}, \bs{z}} \sum_{t=1}^{T} \left(
    y_t - \sum_{p=1}^{P} X_{t,p} \beta_{p} \right)^2
    \label{eq:alg-obj}
    & \quad \text{subject to} \\
    (1-z_p, \beta_{p}) \in \mathcal{SOS}-1,
    & \quad p = 1, \ldots, P, \label{eq:constr-sos}\\
    \sum_{p=1}^P z_p \leq k,
    & \label{eq:constr-gensparse}\\
    \text{s.t.} \ z_p \in \{0,1\},\ \beta_p \in \mathbb{R},
    & \quad p = 1, \ldots,P. \label{eq:alg-zs} 
  \end{align} 
\end{subequations}
The binary variable $z_p$ takes the value 1 if predictor $X_p$ is used in the model and zero otherwise. Special ordered set constraints (\ref{eq:constr-sos}) allow only one of $1-z_p$, or $\beta_p$, to be non-zero. Constraint (\ref{eq:constr-gensparse}) controls the sparsity of the models by restricting the maximum number of predictors to $k$. The value $k$ can be chosen with model selection criteria such as the AIC \citep{akaike1973} or BIC \citep{Schwarz1978}. Alternatively, cross-validation methods can be used (see e.g. \cite{Stone1973}).  The MIQO optimisation problems can be solved efficiently with modern optimisation solvers such as Gurobi \citep{gurobi}. Good feasible solutions can be obtained for models with sparsity $k+1$ using the optimal solution with sparsity $k$. By modifying the right-hand side of constraint (\ref{eq:constr-gensparse}) Gurobi can automatically use the previous optimal solution to `warmstart' the solver.

\paragraph{Treatment of correlated predictors.} \label{sec:correlated}
Similar to \cite{BertsimasKing2016}, we can easily have additional constraints to the MIQO formulation, for example, to avoid including highly correlated predictors within our model. Specifically, we can add the constraint
\begin{equation}
  z_p + z_s \leq 1,
  \ \forall\ (p,s) \in \mathcal{HC}:= \{(p,s) : \text{Cor}(X_p,X_s) > \rho\}. \label{eq:contr-pairwisecorrelation}
\end{equation}
Constraint (\ref{eq:contr-pairwisecorrelation}) allows at most one of $z_p$ or $z_s$ into the model for all pairs of highly correlated variables, specified by the set $\mathcal{HC}$. 
Here, $\rho$ can be seen as the maximum pairwise correlation between predictors that will be permitted to enter a model. 

\paragraph{Incorporating expert knowledge.} \label{sec:expert-knowledge}
In many settings, expert knowledge may suggest predictors that must be present in the model. For example, it may be suitable to account for known outliers or other known external influences. Let the set $\mathcal{J}$ denote the indices of predictors that must be present in the model.  This can be enforced by adding the constraint
\[
  z_p = 1,
  \quad \forall \ p \in \mathcal{J}. 
\]
Expert knowledge may also suggest how the predictors should effect the response variables. For example, some predictors may be known to have a positive effect on the response variables (see e.g.\ Section \ref{sec:data-study}). 
We propose to include this expert knowledge as follows. Let the sets $\mathcal{P}$ and $\mathcal{N}$ denote the sets of predictor indices that should have positive and negative effects on the response variables respectively. Then the constraints
\begin{align*}
  \beta_{p} \geq 0
  \quad \forall p \in \mathcal{P},
  \quad \text{and} \quad 
  \beta_{p} \leq 0
  \quad \forall p \in \mathcal{N},
\end{align*}
ensure that the coefficients take the correct sign according to expert opinion, or the corresponding predictors are excluded from the models.  As well as aiding context-specific interpretability of the models, an additional advantage of enforcing sign constraints is that we have observed that it speeds up the optimisation. 

\paragraph{Including transformations of predictors.}\label{sec:transformations}

In Section \ref{sec:introduction} we discussed the need to determine the best parameter for a set of non-linear transformations of a predictor. To ensure the best parameters are found in terms of minimising the least squares objective, we can use the following constraints. Let $\mathcal{T}_i$ denote the set of predictors obtained by applying a non-linear transformation to a predictor over a grid of values. Then the constraints
\begin{align} \label{eq:constr-nonlin-trans}
  \sum_{p \in \mathcal{T}_i} z_p \leq 1,
  \quad \text{for} \ \mathcal{T}_1, \ldots, \mathcal{T}_I, 
\end{align}
will ensure at most one of the predictors from each group $\mathcal{T}_i$ will appear in the model.

\paragraph{Ensuring model sparsity.}\label{sec:sparsity}
In our motivating application, as well as many other contexts, sparse models are desired to illustrate the strongest effects of a few predictors.   Hence for computational reasons we suggest setting a maximum model sparsity $k_{\text{max}}$. The choice of $k_{\text{max}}$ could be somewhat arbitrary. However, in our formulation the value $k_{\text{max}}$ can be determined automatically by using constraints of the form (\ref{eq:contr-pairwisecorrelation}) and (\ref{eq:constr-nonlin-trans}). These constraints suggests that there exists a maximum level of model sparsity where at least one constraint (\ref{eq:contr-pairwisecorrelation}) or (\ref{eq:constr-nonlin-trans}) will be violated if an additional predictor is included into the model. State-of-the-art optimisation solvers, such as Gurobi \citep{gurobi} will inform the user if an optimisation formulation is infeasible. We propose modifying the sparsity constraint (\ref{eq:constr-gensparse}) as follows:
\[
  \sum_{p=1}^{P} z_p = k.
\]
If $k > k_{\text{max}}$, a feasible solution to the \textit{modified} best subset problem does not exist and the solver will inform the user of an infeasible optimisation model; larger predictor subsets can hence be discounted. 
In practice, an additional choice to reduce computation is to set a maximum runtime of the solver, as suggested by \cite{BerstimasKingMazumder2016}. Often this finds the optimal solution quickly, but may take hours to provide the certificate of optimality.

\section{Simultaneous predictor selection for systems of linear regression models}\label{sec:simult-pred-select}
Interpretability and consistency of models is important in industry. If a model is difficult to interpret then it is of limited use for practitioners trying to understand the dynamics of the system of interest. When models contradict expert opinion or take very different forms for a number of related response variables, the reliability of the models may be questioned. We now describe our proposed extension to the best subset formulation (\ref{eq:univariate-formulation}) to simultaneously select predictors and obtain models for multiple related response variables to ensure consistency in the selected predictors for each response variable.

\subsection{MIQO formulation for multiple response variables}\label{sec:multiple-datasets}

Consider estimating regression models for $M$ response variables, where we assume that these response variables are suitable for joint analysis. We write the system of models as
\begin{equation} \label{eq:response-model-system}
\begin{aligned}
   &Y_{1} = \sum_{p=1}^{P} X_{1,p} \beta_{1,p} + \eta_{1}, \\
   &\ \vdots \hspace{2.5cm} \vdots \\
   &Y_{M} = \sum_{p=1}^{P} X_{M,p} \beta_{M,p} + \eta_{M},
\end{aligned}
\end{equation}
where $\eta_m\sim N(0,\sigma^2_m)$, and $\beta_{m,p} \in \mathbb{R}$
for $p=1,\ldots,P$,\newline $m=1,\ldots,M$.

Here, we assume that each response variable has a unique realisation of the $P$ predictor variables. For example, suppose predictor $X_1$ corresponds to precipitation, then predictor $X_{m,1}$ corresponds to the precipitation for response $Y_m$. Let $\mathcal{S}_m$ denote the set of selected predictors for response $m$. The current procedure used by our industrial collaborator often produces models where $\mathcal{S}_{m_1} \neq \mathcal{S}_{m_2}$, contrary to expert opinion. This motivates the following formulation, which we call the \textit{Simultaneous Best Subset} (SBS) problem: 
\begin{equation}
  \min_{\bs{\beta}} \sum_{m=1}^M \sum_{t=1}^{T} \left(
  y_{m,t} - \sum_{p=1}^{P} X_{m,t,p} \beta_{m,p} \right)^2,
  \label{eq:sbs-prob}\
  \end{equation}
$$\text{subject to} 
  \bigcup_{m=1}^{M} \mathcal{S}_m \leq k.$$
The union $\bigcup_{m=1}^{M} \mathcal{S}_m$ gives the selected predictors across all models: if all models contain the same predictors, then each model may have up to $k$ predictors present.

As well as consistency in predictor selection, some similarity in the coefficients $\beta_{1,p}, \ldots, \beta_{M,p}$ may be expected in considering multiple response variables. We can penalise for large dissimilarities in the coefficients by introducing dummy variables $\bar{\beta}_1, \ldots, \bar{\beta}_P$ and adding the following penalty to the objective appearing in (\ref{eq:sbs-prob}): 
\begin{align}
  \mathcal{P}(\bs{\beta}) = \lambda
  \sum_{m=1}^{M} \sum_{p=1}^{P}
  (\bar{\beta}_p - \beta_{m,p})^2.
  \label{eq:dis-penalty}
\end{align}
The tuning parameter $\lambda$ must be determined. For large $\lambda$ the penalty (\ref{eq:dis-penalty}) will dominate the objective and force the solver to encourage $\beta_{1,p}, \ldots, \beta_{M,p}$ close to $\bar{\beta}_p$, for $p=1,\ldots,P$. In practice, a suitable range of $\lambda$ must be chosen. In what follows, we use a sequence of $\lambda$ values equally spaced on the log scale between $0$ and $2g_{k}$, where $g_{k}$ is the value of the objective of the solution to the SBS problem (\ref{eq:sbs-prob}) with sparsity $k$. 
We have observed that coefficients become more stable for large values of $\lambda$, and that the coefficients $\beta_{1,p}, \ldots, \beta_{M,p}$ become sufficiently close to $\bar{\beta_p}$ for $p=1,\ldots,P$ when $\lambda=2g_{k}$.

The number of binary variables in the optimisation model need not increase when simultaneously estimating multiple regression models -- the number stays at $P$, the number of predictor variables. However, the number of constraints in the optimisation must be increased to ensure a feasible solution of (\ref{eq:sbs-prob}) is obtained. To this end, we use the $\mathcal{SOS}-1$ constraints
\begin{align}\label{eq:SMRMC-sos}
  (1-z_p, \beta_{m,p}) \in \mathcal{SOS}-1, 
  \end{align}
for $p=1,\ldots,P, m=1,\ldots,M$. 
These constraints, along with the sparsity constraint (\ref{eq:constr-gensparse}), ensure that no more than $k$ predictors are present across each of the $M$ regression models. 

Analogous to Section \ref{sec:correlated}, to prevent pairs of highly correlated predictors we define the set of highly correlated predictors $\mathcal{HC}$ in this setting as pairs $(p,s)\in \{1,\ldots,P\} \times \{1,\ldots,P\} $ such that 
\[
\left(\sum_{m=1}^{M} \sum_{p \neq s} \mathbb{1}_{cor(X_{m,p}, X_{m,s}) > \rho }\right) > 0.
\]
By using the constraints of the form (\ref{eq:contr-pairwisecorrelation}), we prevent any model in the system (\ref{eq:response-model-system}) containing pairs of predictors with correlation that exceeds $\rho$.

\subsection{Extension to serially correlated data}
\label{sec:appl-seri-corr} 
Fitting linear regression models to time-ordered data often produces models where the observed residuals appear serially correlated \citep{BrockwellDavis2002}. To remedy this issue, in this section we propose a two-step algorithm, similar in spirit to that of \cite{CochraneOrcutt1949} that implements a predictor selection step to a generalised least squares (GLS) transform of the data. In what follows, we give an example of the GLS transform, before describing how we incorporate predictor selection. 

Suppose we have a response variable $Y$ and predictors $X_1, \ldots, X_P$, and suppose the true model for the relationship between the response and predictors is 
\begin{subequations}
\begin{alignat}{1}
  Y_t &= \sum_{p=1}^{P} X_{t,p} \beta_P + \eta_t
  \label{eq:example-regsarima-reg}
  \quad \text{where} \\
  \eta_t &= \phi \eta_{t-1} + e_{t}.
  \label{eq:example-regsarima-sarima}
\end{alignat} \label{eq:example-regsarima}
\end{subequations}
In this setting, the regression residuals $\eta_t$ are serially correlated. Ignoring serial correlation in observed residuals not only mis-specifies the model but ignores potentially valuable information. Minimising the least squares objective (\ref{eq:OLS}) no longer gives the most efficient estimator for the regression coefficients \citep{RaoToutenburg-LinearModels}. Providing (\ref{eq:example-regsarima-sarima}) is stationary \citep[see][]{BrockwellDavis2002} we can write (\ref{eq:example-regsarima}) as a regression model with residuals that are {\em not} serially correlated via
\begin{align}
  \frac{Y_t}{1 - \phi L} = \sum_{p=1}^{P} \frac{X_{t,p}}{1 - \phi L} \beta_p + e_t,
  \label{eq:twostep-ex-filtered}
\end{align}
where $L$ denotes the backward-shift operator such that $L \eta_t = \eta_{t-1}$. The linear filter can be applied to the response and predictor variables to obtain transformations of the original variables.  In other words, the original variables can be written $\tilde{Y}_t = \frac{Y_t}{1 - \phi L}$ and $\tilde{X}_{t,p} = \frac{X_{t,p}}{1 - \phi L}$. We show empirically in Section \ref{sec:sim-appl-seri-corr} that predictor selection accuracy can be improved by transforming the response and predictor variables appropriately. 

In general, neither the predictor variables present in the model or the serial correlation structure of the regression residuals are known. We assume a general regression seasonal autoregressive integrated moving average (Reg-SARIMA) model of the form
\begin{subequations} \label{eq:reg-SARIMA}
  \begin{equation} \label{eq:r-SARIMA-reg}
    Y_{m,t} = \sum_{p=1}^{P} X_{m,p,t} \beta_{m,p} + \eta_{m,t}, 
  \end{equation}
  where
  \begin{equation} \label{eq:r-SARIMA-SARIMA}
    \eta_{m,t} = \frac{\theta_m(L) \Theta_m(L^s)}{\nabla^{d_m} \nabla_s^{D_m} \phi_m(L) \Phi_m(L^s)} \epsilon_{m,t},
  \end{equation}
\end{subequations}
and propose the following two-step algorithm to determine the best predictors and autocorrelation structure of the regression residuals. First we seek suitable predictors for the model. We fix the sparsity $k$ and use the data
$
  (Y_1, X_{1,1}, \ldots, X_{1,P}), \ldots, (Y_M, X_{M,1}, \ldots, X_{M,P})
  $
  to determine a suitable set of predictors by solving the SBS problem. Given initial estimates of the coefficients $\hat{\beta}_{1,1}^{k,0}, \ldots, \hat{\beta}_{M,P}^{k,0}$, we then obtain the observed residuals for each model
\[
  \hat{\eta}^{k,0}_{m,t} = y_{m,t} - \sum_{p=1}^{P} X_{m,p,t} \hat{\beta}_{m,p}^{k,0}.
\]
We need to estimate the serial correlation structure of the regression residuals. Given a list $\mathcal{L}$ of suitable SARIMA models, these models can be fit to the observed regression residuals $\hat{\eta}^{k,0}_{m,t}$ for $m=1, \ldots, M$ and the best SARIMA model identified for each $m$, for example, based on an appropriate information criterion. We require the transformed data
\begin{eqnarray} \label{eq:two-step-filtered}
  \frac
  {\nabla^{\hat{d}_m} \nabla_s^{\hat{D}_m} \hat{\phi}_m(L) \hat{\Phi}_m(L^s)}
  {\hat{\theta}_m(L) \hat{\Theta}_m(L^s)}
  Y_{m,t} &=& \tilde{Y}_{m,t} \quad \text{and} \\
  \label{eq:two-step-filtered2}
  \frac
  {\nabla^{\hat{d}_m} \nabla_s^{\hat{D}_m} \hat{\phi}_m(L) \hat{\Phi}_m(L^s)}
  {\hat{\theta}_m(L) \hat{\Theta}_m(L^s)}
  X_{m,p,t} &=& \tilde{X}_{m,p,t},
\end{eqnarray}
for $m=1,\ldots,M.$ 

Consider fitting the SARIMA model (\ref{eq:r-SARIMA-SARIMA}) to obtain the observed model errors $\hat{\epsilon}_{m,t}$,
\[
  \hat{\eta}_{m,t}
  \frac
  {\nabla^{\hat{d}_m} \nabla_s^{\hat{D}_m} 
    \hat{\phi}_m(L) \hat{\Phi}_m(L^s)}
  {\hat{\theta}_m(L) \hat{\Theta}_m(L^s)}
  =
  \hat{\epsilon}_{m,t}.
\]
This process can be applied to (\ref{eq:two-step-filtered}) and (\ref{eq:two-step-filtered2}) to obtain $\tilde{Y}_{m,t}$ and $\tilde{X}_{m,p,t}$ for $m=1\ldots,M$ and $p=1,\ldots,P$. Lastly, the predictors can be re-selected by solving the SBS problem again with the filtered data, $\tilde{Y}_{m,t}$ and $\tilde{X}_{m,p,t}$. This procedure can be iterated until convergence in the regression estimates, selected predictors, and the models for serial correlation. If the procedure does not converge quickly an upper limit to the number of iterations can also be considered. However, we have observed that convergence often occurs after two iterations. The pseudo-code for our two-step procedure, {\em Two-stage Simultaneous Predictor Selection} ({\tt SPS2}) is shown in given in Algorithm 
\ref{fig:two-step_alg}.
  \begin{algorithm}
  \caption{Pseudo-code for the two-step subset selection algorithm ({\tt SPS2}) allowing for serial correlation. \label{fig:two-step_alg}} 

    \SetAlgoLined
    \KwResult{}
    Input: $\bs{Y}, \bs{X}$ and $\mathcal{L}$ \;

    \For{k in $\{1, \ldots, P\}$}{
      Apply SBS to $(Y_1, X_{1,1}, \ldots, X_{1,P}), \ldots, (Y_M, X_{M,1}, \ldots, X_{M,P})$;

      Obtain $\hat{\bs{\beta}}^{k,1}$
      
      \For{$(p,d,q)(P,D,Q,s) \ \text{in} \ \mathcal{L}$}
      {
        \For{$m$ in $\{1,\ldots,M\}$}
        {
          Fit SARIMA$(p,d,q)(P,D,Q,s)$ to $\hat{\eta}^{k,1}_{m}$

          Select best $(p,d,q)(P,D,Q,s)$ giving $\hat{\phi}_m^{k,1}, \hat{\Phi}_m^{k,1}, \hat{\theta}_m^{k,1}, \hat{\Theta}_m^{k,1}$
        }
      }
      Filter $\bs{Y}, \bs{X}$ giving $\tilde{\bs{Y}}^{k,1}, \tilde{\bs{X}}^{k,1}$
      
      \For{$it$ in $\{1, \ldots, maxiter\}$}
      {
        Apply SBS to $\tilde{\bs{Y}}^{k,it-1}, \tilde{\bs{X}}^{k,it-1}$;

        Obtain $\hat{\bs{\beta}}^{k,it-1}$

        \For{$(p,d,q)(P,D,Q,s) \ \text{in} \ \mathcal{L}$}
        {
          \For{$m$ in $\{1,\ldots,M\}$}
          {
            Fit SARIMA$(p,d,q)(P,D,Q,s)$ to $\hat{\eta}^{k,it-1}_{m}$

            Select best $(p,d,q)(P,D,Q,s)$ giving $\hat{\phi}_m^{k,it-1}, \hat{\Phi}_m^{k,it-1}, \hat{\theta}_m^{k,it-1}, \hat{\Theta}_m^{k,it-1}$
          }
        }
        \If{converged}{Return}
      }
    }
    \bigskip
  \end{algorithm}

\section{Performance on simulated data} \label{sec:simulation-study}
In this section we investigate the properties of our simultaneous predictor selection approach. In particular, we perform a number of simulations investigating how our SBS model compares to applying the standard best subset approach to estimate each linear regression model separately. We compare our simultaneous estimation procedure to the LASSO \citep{Tibshirani1996} and elastic net \citep{ZouHastie2005_ElasticNet}. We also compare our approach to an alternative simultaneous estimation procedure: we modify the Simultaneous Variable Selection approach of \cite{TurlachVenablesWright2005} to estimate the system of linear models (\ref{eq:response-model-system}). We call this approach \texttt{SVS-m}, the {\em modified} SVS approach; further algorithmic details of this procedure can be found in Appendix \ref{sec:addit-algor-deta}.

We generate data from Model (\ref{eq:response-model-system}) where we fix the regression coefficients as
\begin{align*}
  \beta_{m,p} =
  \begin{cases}
    0.3, \quad & \text{for} \ p = 17, \\
    1, \quad & \text{for} \ p = 18, \\
    0.6, \quad & \text{for} \ p = 19, \\
    0, \quad & \text{otherwise}, 
  \end{cases} \quad \text{for all} \ m.
\end{align*}

The predictors and residuals are simulated as follows:
\begin{eqnarray}\label{eq:simspec}
  \bs{X}_{m,t} \sim \text{MVN}_{35}(\bs{0}, \bs{\Sigma}_{x}), \quad
  \eta_{m,t} \sim \text{N}(0, \sigma_{\eta}^2),\\ \text{where} \ \nonumber
  \bs{\Sigma}_{x} := (\bs{\Sigma}_{x})_{i,j} = \rho^{|i-j|}. \quad
\end{eqnarray}

The particular values of the residual variance, $\sigma_{\eta}^2$ and predictor correlation, $\rho$ will be clarified in each simulation. When the correlation between predictors is large, the predictors $X_{17}$, $X_{18}$, and $X_{19}$ become hard to distinguish and hence accurately selecting the correct generating predictors is challenging. We use $P=35$ predictor variables as provably optimal solutions can be obtained within seconds for sparse models (see Appendix \ref{app:simstudy}). 

In the simulations that follow we solve the SBS problem with $M=1,5,10,20,35$, increasing the number of regression models used for simultaneous predictor selection and coefficient estimation. Note that $M=1$ corresponds to the best subset approach of \cite{Miller2002}. In a simulation of size $N$, we record the number of times each application of the SBS approach recovers the true subset by applying the SBS approach with the sparsity set to the true value, $k=3$. We also record the mean squared error of the regression coefficients given by
\[
  \text{MSE}(\bs{\beta}) = \frac{1}{MP} \sum_{m=1}^{M} \sum_{p=1}^{P} \left( \beta_{m,p} - \hat{\beta}_{m,p} \right)^2,
\]
where $\hat{\beta}_{m,p}$ is the estimate of $\beta_{m,p}$. This measure will penalise large deviations from the true coefficients and take account of potential variation as we change the value of $M$. Unless specified otherwise, we do not apply shrinkage as we wish to demonstrate the gains from simultaneous selection only. 

\subsection{Evaluation of simultaneous predictor selection} \label{sec:joint-modelling}

\subsubsection{Effect of correlated predictors}

We start by investigating how predictor correlation affects selection accuracy for the best subset method, and how this improves for simultaneous predictor selection as the number of jointly-estimated models increases. We generate $N=1000$ synthetic datasets using the specification \eqref{eq:simspec} and fix $\sigma_{\eta}^2 = 1$.
\begin{figure*}[!h] \centering 
  \begin{subfigure}{.49\linewidth}
    \includegraphics[width=0.95\linewidth]{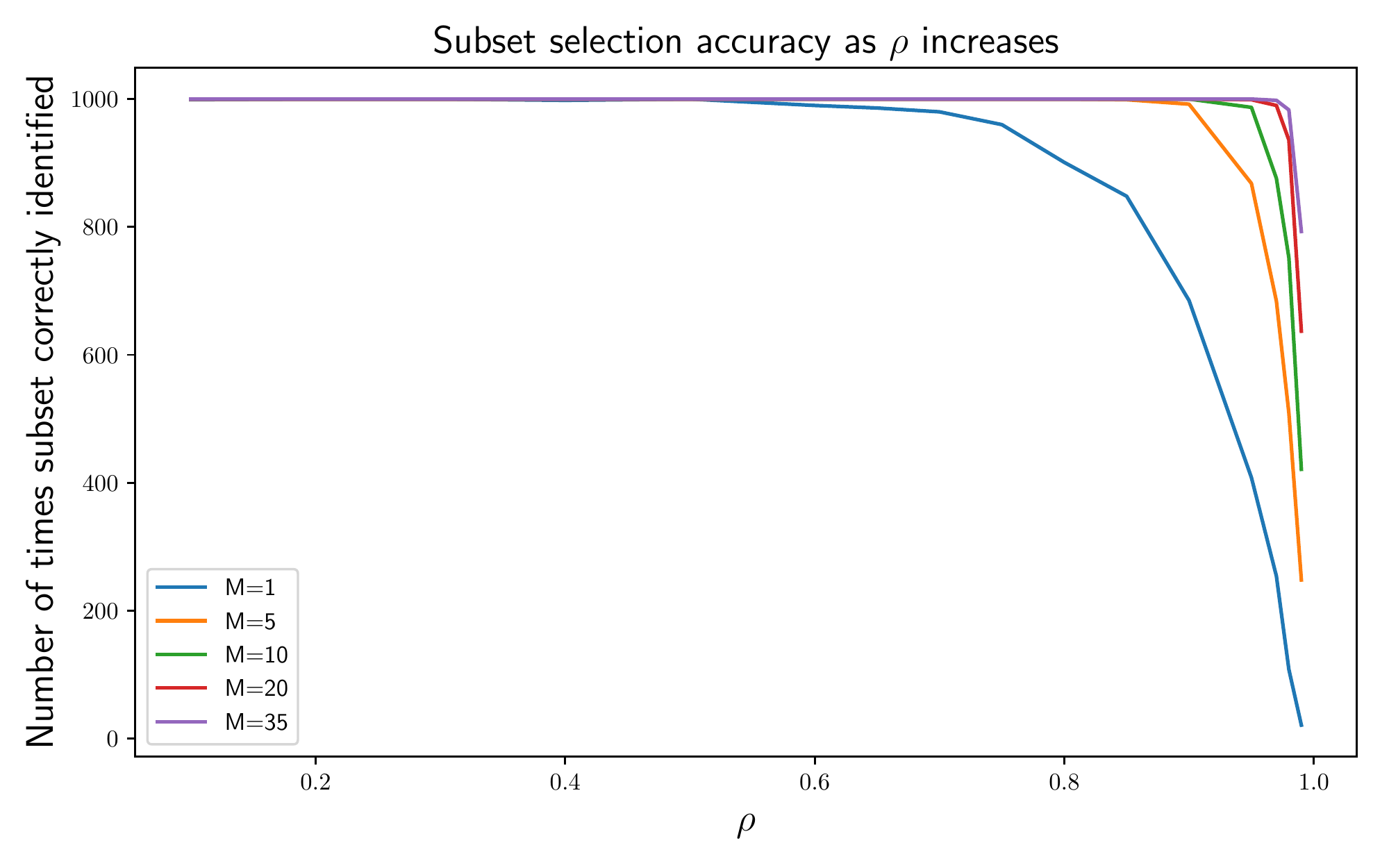}\caption{}
    \label{fig:numcorrect-increasing-M-rho}
  \end{subfigure}\hfill
  \begin{subfigure}{.49\linewidth}
    \includegraphics[width=0.95\linewidth]{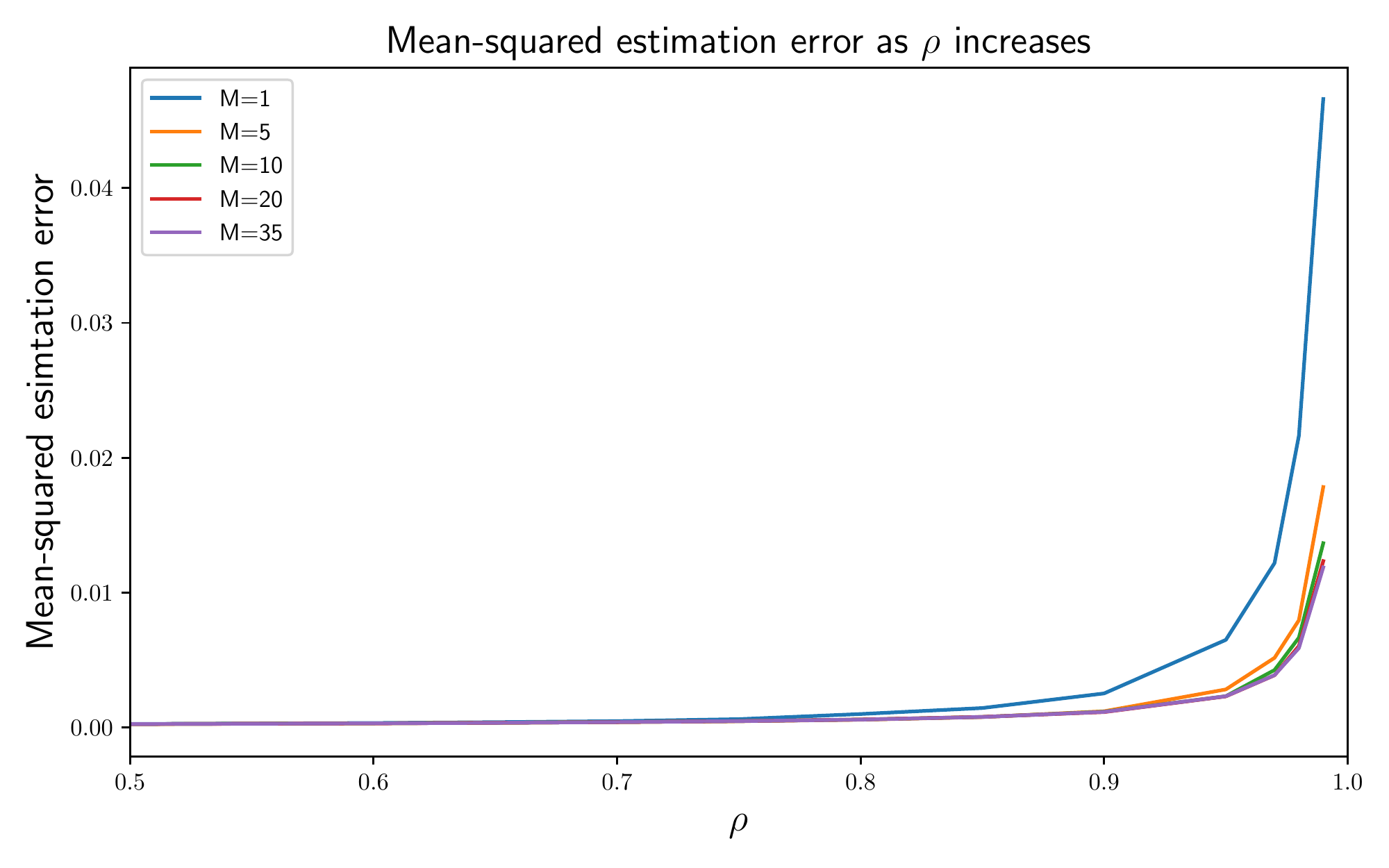}\caption{}
    \label{fig:error-beta-increasing-M-rho}
  \end{subfigure}
  \caption{Performance of SBS as $\rho$, the correlation between predictors, increases for different numbers of jointly-modelled response variables, $M$: (a) Selection accuracy; (b) mean squared error of regression coefficient estimates.}
\end{figure*}

Figure \ref{fig:numcorrect-increasing-M-rho} shows the selection accuracy for simultaneous subset selection with differing values of $M$.   We see that for the best subset method ($M=1$), the accuracy deteriorates rapidly as the predictor correlation, $\rho$, exceeds 0.5. However, simultaneous predictor selection increases the correlation threshold at which selection accuracy deteriorates to 0.87 with just five models. Consequently, the mean squared error in coefficient estimates decreases, as can be seen from Figure \ref{fig:error-beta-increasing-M-rho}. Selection accuracy is seen to improve further with a greater number of models estimated simultaneously. 

We also investigate the performance of SBS with increasing residual variance, $\sigma^2_{\eta}$ for differing values of $M$ and data length, $T$; as one might expect, with increasing residual variance it is much harder to recover the true predictors.  For reasons of brevity, these results are deferred to Appendix \ref{app:simstudy}.

\subsubsection{Simultaneous shrinkage}\label{sec:simult-shrink}
The coefficients obtained from minimising the least squares objective with highly correlated predictors can suffer from high variance. As such, the variation in selected predictors for the best subset method is also high, as shown in Section \ref{sec:multiple-datasets}, mirroring the observations by \cite{HastieTibshiraniFriedmanESL}. To investigate the effect of shrinking coefficients for each predictor towards a common value, we fix $M = 5$ and simulate $T=750$ observations for each response variable and their associated predictors from the model \eqref{eq:simspec}. We split the data randomly into two sets, using 500 observations for each response
variable as a training set to estimate the models. The remaining 250 observations are used to determine the predictive accuracy of the models. We fix $\rho=0.95$ and $\sigma_{\eta}^2 = 2$ and again consider when $k=3$.  to show the effects on in-sample and out-of-sample prediction error.  
\begin{figure*}[!h]
  \centering 
  \begin{subfigure}{.49\linewidth}
    \includegraphics[width=0.95\linewidth]
    {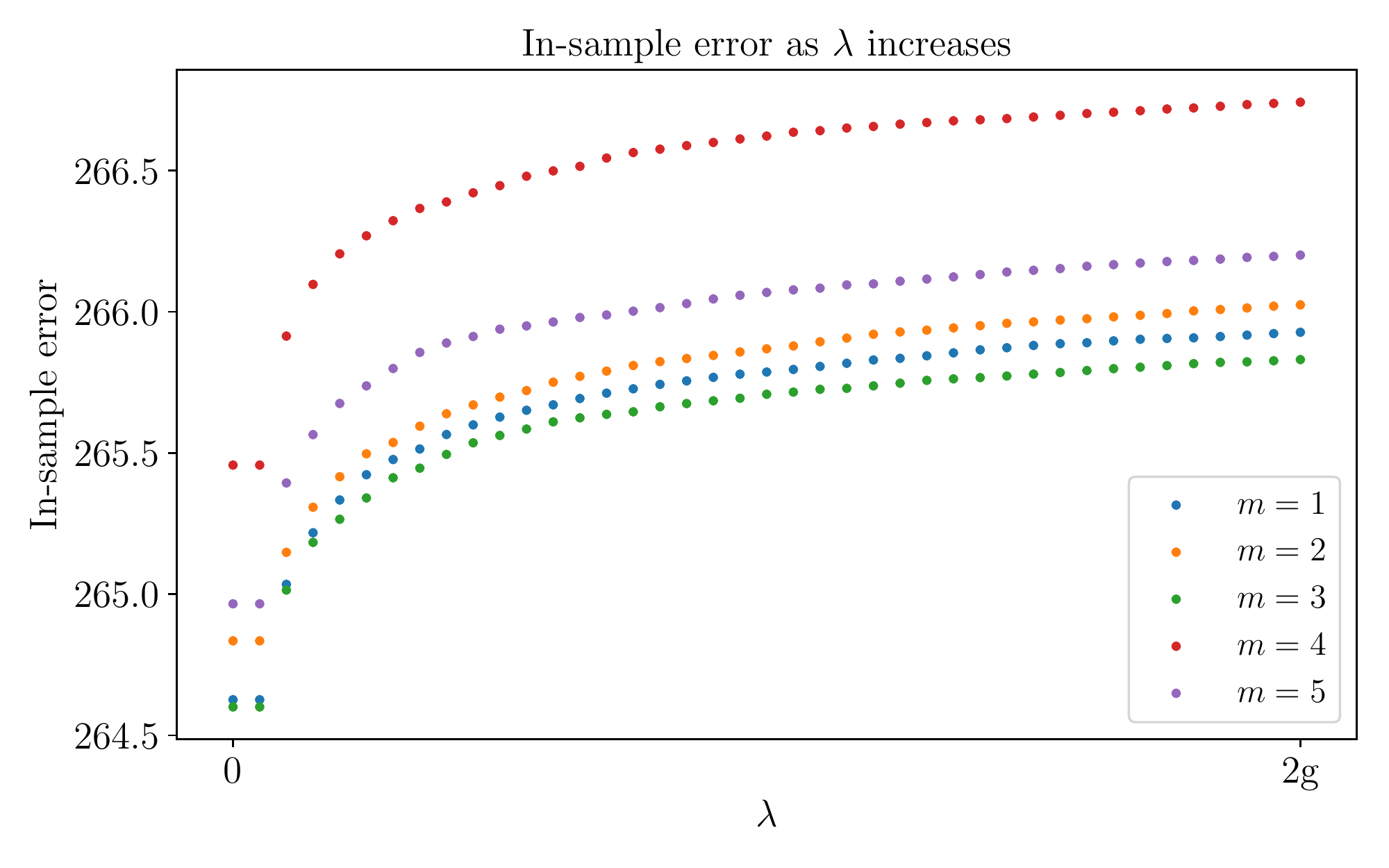}\caption{}
    \label{fig:multivariate-shrinkage-insample-error}
  \end{subfigure}
  \begin{subfigure}{.49\linewidth}
    \includegraphics[width=0.95\linewidth]
    {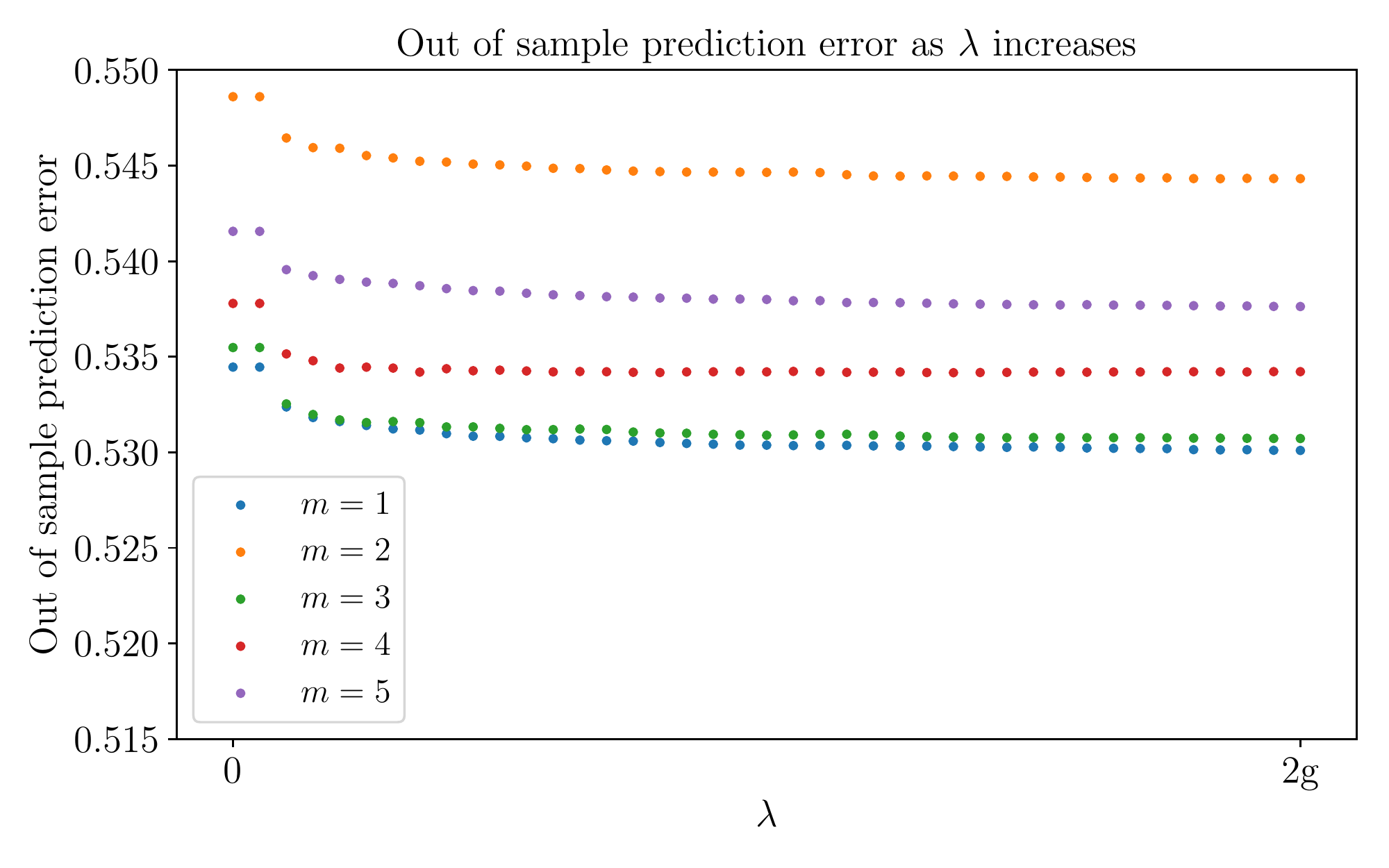}\caption{}
    \label{fig:multivariate-shrinkage-outsample-error}
  \end{subfigure}
  \caption{Mean squared error when coefficient shrinkage across the $M$ regression models is imposed: (a) in-sample error; (b) out-of-sample error.\label{fig:multivariate-shrinkage-error}}
\end{figure*} 

Figure \ref{fig:multivariate-shrinkage-error} shows the MSE for both scenarios over a range of increasing penalty values, $\lambda$. By penalising the differences in $\beta_{1,p}, \ldots, \beta_{M,p}$ for $p=1,\ldots,P$, we bias the estimates of the regression coefficients, increasing the in-sample error (see Figure \ref{fig:multivariate-shrinkage-insample-error}).  However this leads to improved out-of-sample prediction error (see Figure \ref{fig:multivariate-shrinkage-outsample-error}) as information is shared across regression models by shrinking the coefficients for each predictor to a common value. 
 
\begin{figure*}[!h]
  \centering
  \includegraphics[width=0.95\linewidth]
  {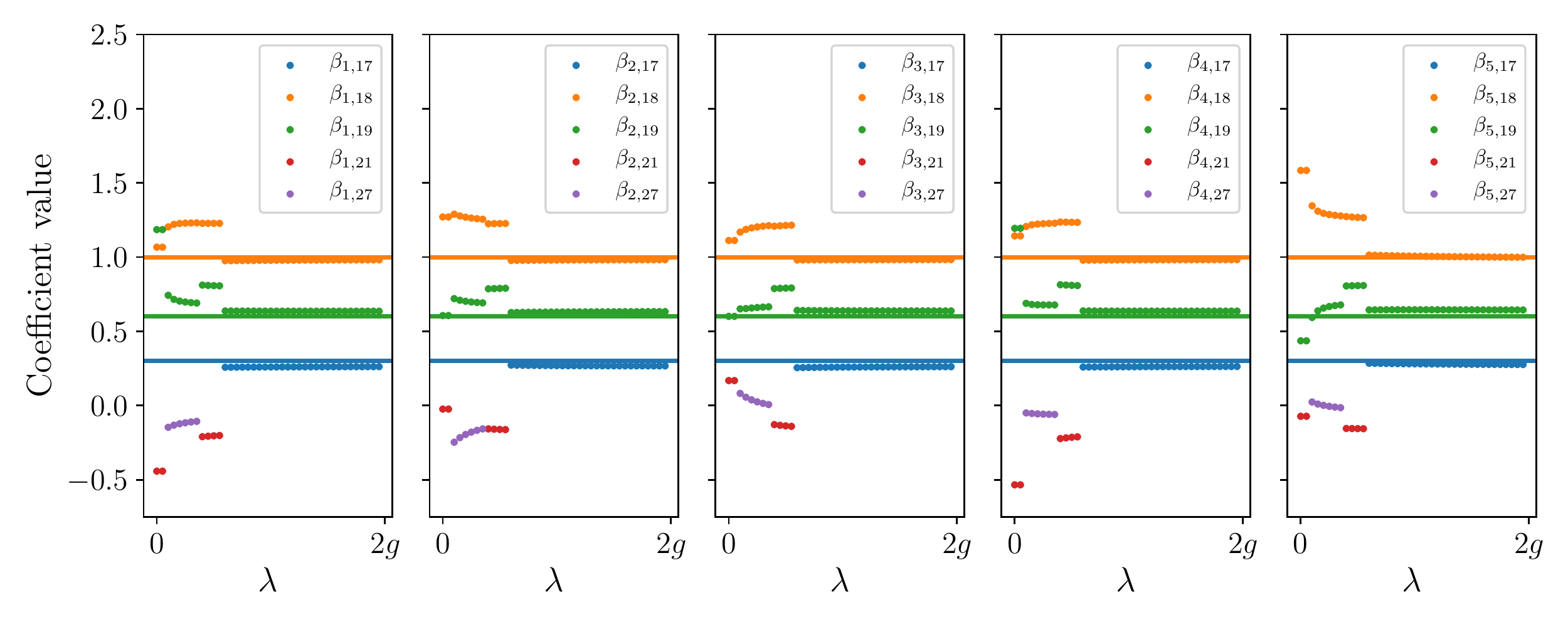}
  \caption{Trace plot of the regression coefficients $\bs{\beta}_1, \ldots, \bs{\beta}_5$ (from left to right), as the shrinkage parameter $\lambda$ is increased, penalising dissimilarities in the coefficients.}
  \label{fig:multivariate-shrinkage-traceplot-beta}
\end{figure*}

Figure \ref{fig:multivariate-shrinkage-traceplot-beta} shows shows trace plots of the regression coefficients (for one simulated dataset) for each of the five response variables in the system, as the value of the simultaneous shrinkage penalty increases. The horizontal lines show the coefficients of predictors $X_{17}$,$X_{18}$, and $X_{19}$. 

As the penalty increases, the simultaneous best subset changes. 
Despite seeking the best subset of predictors given the true level of sparsity, the true predictors are not initially selected upon solving the SBS problem. Two of the three predictors are correctly identified although the estimates for each model are rather far from the truth. A spurios (zero) predictor is also selected with relatively large coefficients for some of the models (indicated by non-zero coefficients for $\beta_{m,21}$ and $\beta_{m,27}$). As the strength of the joint shrinkage is increased, the noisy predictor leaves for the true third predictor, re-enters the models, upon being replaced finally for the true third predictor again. At this point, the coefficients for all three predictors in each of the regression models appear significantly closer to the true values in comparison to the solutions obtained upon solving the initial SBS problem. 

\subsection{Performance on serially correlated data}
\label{sec:sim-appl-seri-corr}
In Section \ref{sec:appl-seri-corr} we motivated the need to consider autocorrelated regression residuals in predictor selection problems. In this section we demonstrate that we can recover both the true predictors and correlation structure of the regression residuals using the two-step algorithm described in Section \ref{sec:appl-seri-corr}. To this end, we simulate data from Model (\ref{eq:response-model-system}) but now impose a correlation structure on the residuals, taking the form  
\begin{align}\label{eq:sarima-order}
  \eta_{m,t} = 0.9\, \eta_{m,t-1} + e_{m,t} \quad \text{for} \ m = 1,\ldots,5, 
\end{align}
with $e_{m,t}\sim N(0,1)$, i.e.\ the residuals $\eta_{m,t}$ follow an AR(1) or SARIMA(1,0,0)(0,0,0,0) model. The predictors and regression coefficients are the same as those in Section \ref{sec:multiple-datasets}. Our industrial collaborator often observes large changes in the predictors that are selected when the number of observations available changes only slightly. For $N=50$ datasets of length $T=600$ simulated  under model \eqref{eq:sarima-order}, we apply our two-step algorithm (with $k=3$) to each simulated dataset a total of six times: first we use the first 500 datapoints, then first 520 and so on until all 600 points are used. 

\begin{figure*}[!h]
  \centering
  \begin{subfigure}{0.95\linewidth}
    \includegraphics[width=\linewidth]
    {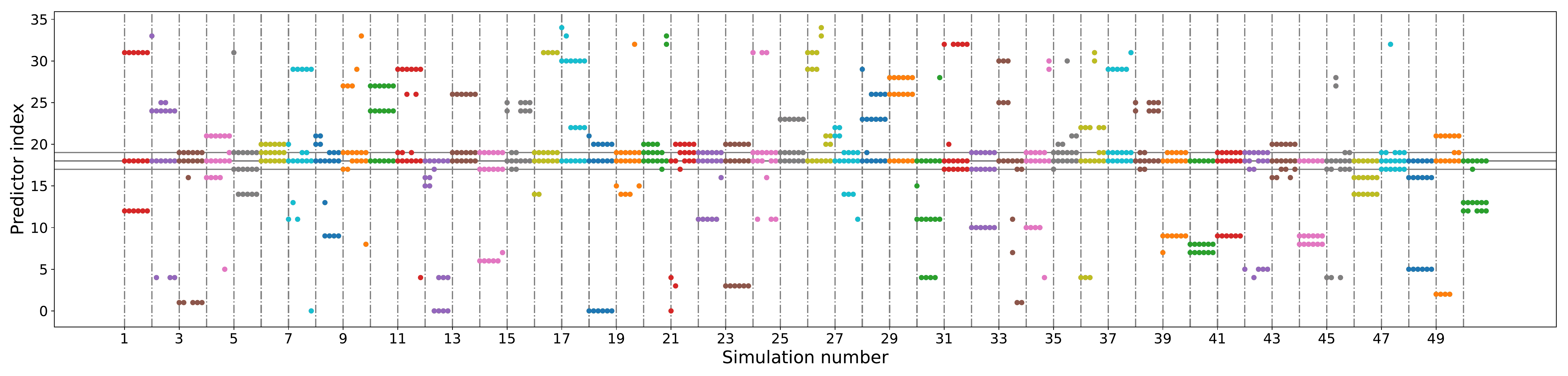}\caption{}
    \label{fig:traceplot-selected-predictors-onestep}
  \end{subfigure}
  \begin{subfigure}{0.95\linewidth}
    \includegraphics[width=\linewidth]
    {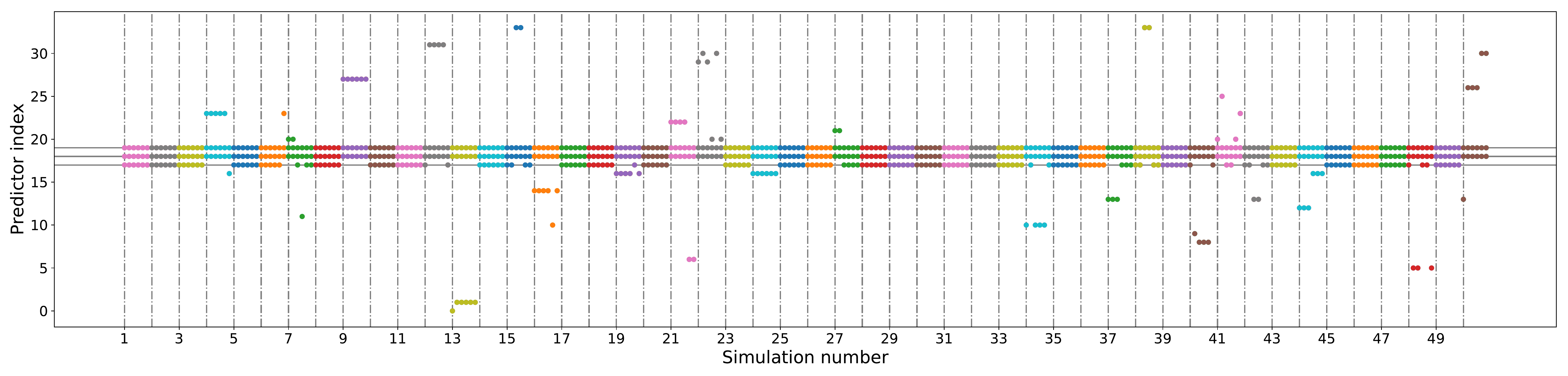}\caption{}
    \label{fig:traceplot-selected-predictors-twostep}
  \end{subfigure}
  \caption{Comparison of the predictors selected using the standard approach and two-step iterative approach: (a) standard procedure, ignoring autocorrelation in the regression residuals (unfiltered covariate selection); (b) two-step procedure {\tt SPS2} (filtered covariate selection).}
  \label{fig:traceplot-selected-predictors}
\end{figure*}

We highlight the predictors selected in each application with and without using the two-step algorithm in Figure \ref{fig:traceplot-selected-predictors}. The selected predictors for each of the simulated datasets are shown within each set of vertical lines. From left to right the vertical triplet of dots indicate the selected predictors for $T=500, 520, \ldots, 600$ within each set of vertical bars. 

For the standard selection procedure, the variation of selected predictors within each dataset is quite alarming as well as the range of predictors across different simulated datasets, reflecting the sensitivity to data length as experienced by our industrial collaborator. This is shown in Figure~\ref{fig:traceplot-selected-predictors-onestep}. In comparison, using the two-step algorithm (Figure \ref{fig:traceplot-selected-predictors-twostep}) we observe much less variation in the selected predictors. Further, the algorithm selects the true predictors in many cases.

\begin{figure*}[!h]
  \centering
  \includegraphics[width=0.97\linewidth]
  {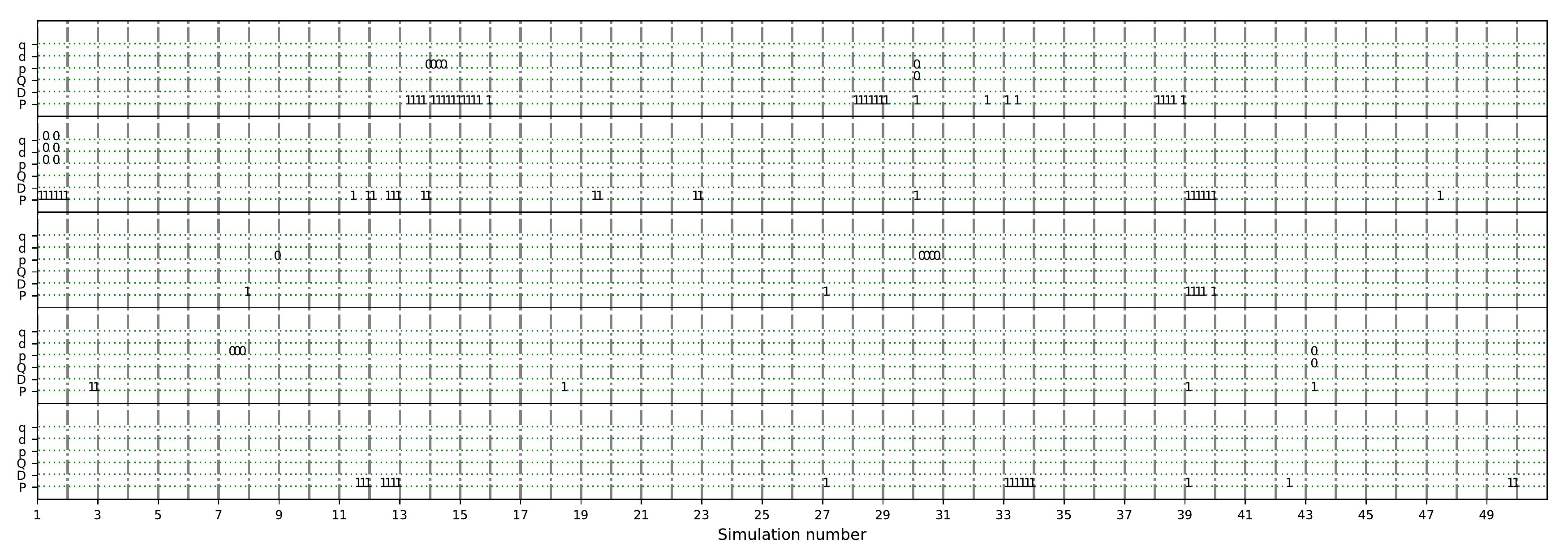}
  \caption{Indication if the true SARIMA model order was identified by the {\tt SPS2} algorithm for each of $N=50$ datasets simulated from model \eqref{eq:sarima-order}.}
  \label{fig:identify-tsorder-trace}
\end{figure*}

We now investigate how well we can recover the true correlation structure of the regression residuals. Recall that the correct model order from specification (\ref{eq:sarima-order}) is $(1,0,0)(0,0,0)$. Figure \ref{fig:identify-tsorder-trace} shows that the model order was correctly identified for a particular simulation if `.' appears on each row, or the value of the order $(p,d,q)(P,D,Q)$ chosen if it were mis-specified.

From Figure \ref{fig:identify-tsorder-trace}, we see that correct values were chosen for the majority of values of the six model orders $(p,d,q)$ and $(P,D,Q)$.  We observe that at least one autoregressive parameter was used ($p \geq 1$) for each dataset, sometimes erroneously using more or including another term, however this is often the case with model selection criteria such as the AIC or BIC.  Modifying the penalty used to select the regression residual model may improve accuracy of selecting these models.

\subsection{Comparison to other approaches}\label{sec:comp-other-appr}
In this simulation we replicate the scenario that motivated our SBS approach. In particular, we simulate series with five blocks of highly correlated predictors. A block of predictors is denoted $\bs{X}_{(b)} = [X_{(b,1)}, \ldots, X_{(b,N_b)}]$. The predictors are simulated as 
\[
  \bs{X}_{(b)} \sim \text{MVN}_{b+4}(\bs{0}, \Sigma_{(b)}),
  \ \text{with}\
  {\Sigma_{(b)}}_{i,j} := 0.95^{|i-j|},
\]
for $b=1,\ldots, 5$. We vary the positions of the active predictors relative to their blocks and the values of the regression coefficients. The regression coefficients take the form
\[
  \beta_{m,p} =
  \begin{cases}
    1, \quad \text{if} \ p = 30, \\
    0.775, \quad \text{if} \ p = 25, \\
    0.55,  \quad \text{if} \ p = 14, \\
    0.325, \quad \text{if} \ p = 5,  \\
    0.1, \quad \text{if} \ p = 2, \\
    0, \quad \text{otherwise}
  \end{cases} \quad \text{for} \ m=1,\ldots,5.
\]
Our primary goal is to compare SBS to current methods in the literature. We apply the elastic net using the \texttt{glmnet} package \citep{elastnet} implemented in the \texttt{R} statistical software \citep{R}, over the default values, $\alpha=0,0.01,\ldots,1$ and for 100 values of $\lambda$ to produce a model for each $m=1,\ldots,5$. We train each model with $T=500$ observations and then use the mean squared prediction error
\begin{equation}\label{eq:mean-square-pred-error}
  \frac{1}{T} \sum_{t=1}^{T}
  \left(
    y_{m,t} - \sum_{p=1}^{P} X_{m,p,t} \hat{\beta}_{m,p}
  \right)
\end{equation}on a 250 observation held-out test set to select the best elastic net model for each $m=1,\ldots,5$. We also compare our results to a forward stepwise algorithm using the standard \texttt{step} function \citep{R} for each $m$, selecting the best model by AIC. We also apply the modified SVS approach ({\tt SVS-m}), as well as a variant  with the regression coefficients constrained to be positive which we denote $\mathtt{SVS-m^+}$. We select the models fit by the simultaneous procedures by considering the simultaneous mean squared prediction error defined in (\ref{eq:mean-square-pred-error}). 

For each of the selected models we record the following performance measures averaged over $N=50$ datasets across each of the models for the $M$ response variables:
\begin{itemize}
\item The average number of predictors (model sparsity), $\hat{k} = \sum_{m=1}^{M} \sum_{p=1}^{P}\mathbb{1}_{\beta_{m,p} \neq 0}$.
\item The mean squared prediction error on a 250 observation held-out validation set.
\item The number of models containing the true subset of predictors. 
\item The number of models that included at least one negative coefficient.
\end{itemize}

The average model sparsity will help inform the interpretability of the models, whilst the prediction error allows us to compare the performance numerically. The average number of models containing the true subset indicates the accuracy of each method as a predictor selector. By counting the number of models with negative coefficients, we can compare how often our industrial collaborator may have obtained misleading models. Note that the elastic net uses 100 values of both $\alpha$ and $\lambda$ which fits 1000 elastic net models to each response variable. 

\begin{table*}[!h]
\centering
  \begin{tabular}{lrrrr}
    \hline
    & Average Sparsity & Average MSE & True Subset
    & Negative Coefficients\\
    \hline
    \texttt{SBS}
    & \textbf{4.70} & \textbf{9.103} & \textbf{0.5}
    & \textbf{0.00}\\
    \texttt{SVS-m}
    & 17.50 & 9.204 & 0.0
    & 1.00\\
    $\mathtt{SVS}$-$\mathtt{m^+}$
    & 12.10 & 9.172 & 0.0
    & \textbf{0.00}\\
    \texttt{Step-f}
    & 9.14 & 10.870 & 0.0
    & 0.92\\
    $\mathtt{enet} (\alpha=1)$
    & 15.04 & 9.358 & 0.0
    & 0.92\\
    \hline
  \end{tabular}
  \caption{Comparative performance of the predictor selection algorithms using the measures described in the text.}
  \label{tab:summarymeasures}
\end{table*} 

The summary measures of all approaches are shown in Table \ref{tab:summarymeasures}. Our proposed SBS approach produces the sparsest models aided by the transformation constraints (\ref{eq:constr-nonlin-trans}), with the average sparsity being slightly lower than the true sparsity. The most likely cause of this is due to not selecting predictor 2 (the relative value of the coefficient is small in comparison to the other predictors). The only method able to recover the true subset was our SBS approach in half of the simulations. The SBS and SVS-m$^+$ techniques always include coefficients with positive values and were the only approaches which did so. All other methods included at least one negative coefficient in a high number of models. 

\begin{figure*}[!h]
  \centering
  \includegraphics[width=0.7\linewidth]{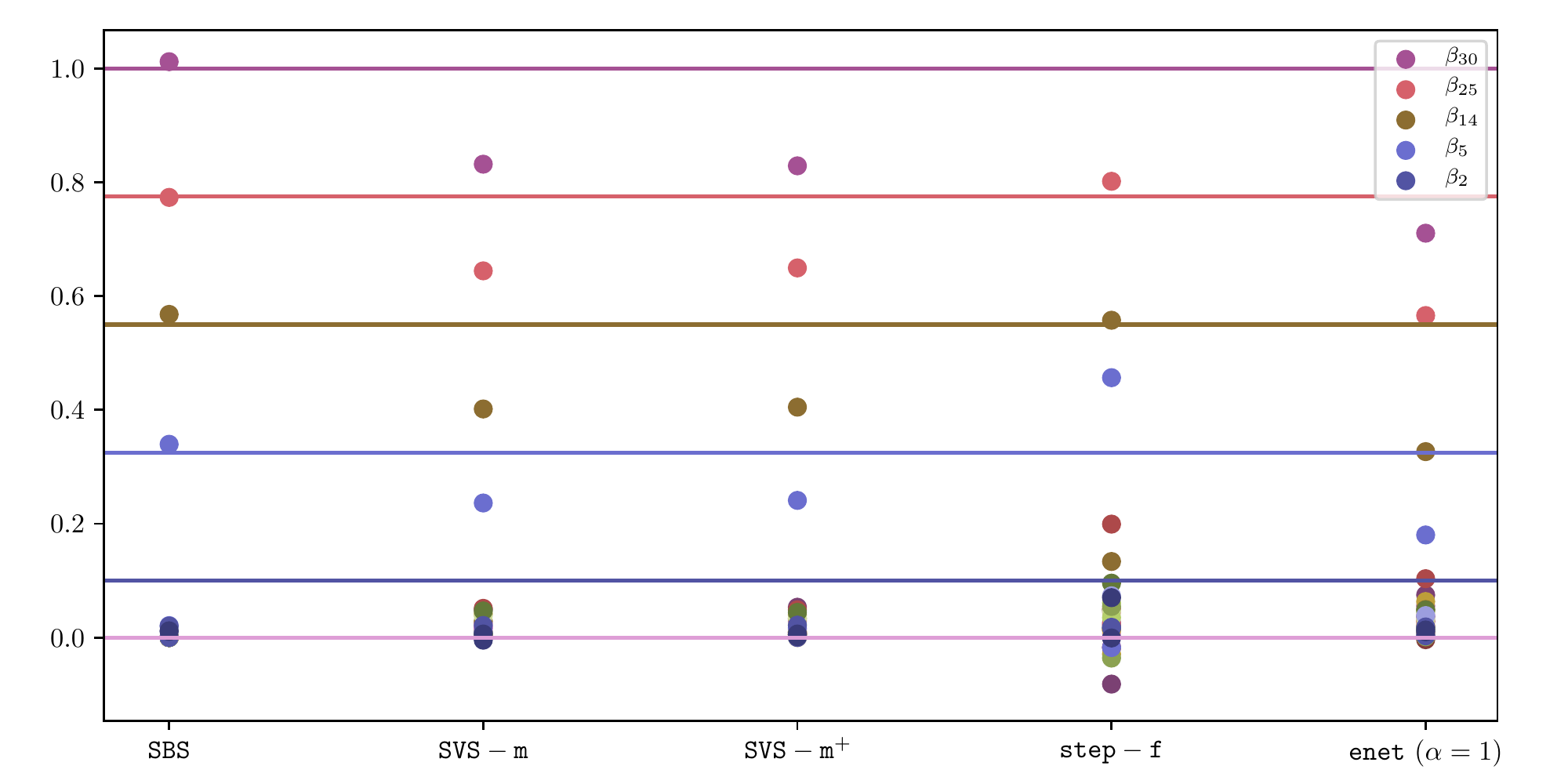}
  \caption{Average estimate of the regression coefficients for each of the methods considered.} \label{fig:avgbeta}
\end{figure*}

Figure \ref{fig:avgbeta} shows the average estimate for the regression coefficients for each of the methods in the study. With the exception of predictor 2, the SBS method appears to give unbiased estimates. Underestimating $\beta_{m,2}$ is likely caused by the small coefficient value where the predictor was not included. The other methods tend to underestimate all of the coefficients which may be expected since they are all shrunk towards zero.

We have also investigated computational aspects (e.g. runtime) of our SBS approach when varying the number of response variables, $M$.  For reasons of brevity we do not include this here, but further details can also be found in Appendix \ref{app:simstudy}.

\section{Telecommunications data study} \label{sec:data-study}

We now demonstrate our proposed methodology on a dataset provided by our industrial collaborator.  In our motivating application, the total number of daily events in a telecommunications network are recorded by type and location within the network. Each type of event may be influenced by a different set of predictors.  For the dataset we consider here, location corresponds to a geographic location, but more detailed information such as the location within the network is available in other applications. We use three response variables of the same type (denoted R1, R2 and R3) from regions in the network considered to be suitable for joint modelling. Urban or rural classifications may help determine if response variables are suitable for joint modelling.  There are a total of 1396 daily observations, corresponding to about 3 years 9 months of data.

We use five groups of predictor variables. Motivated by the remarks in Section \ref{sec:transformations} in relation to weather variables, the first four groups of predictors are derived from transformations applied to the following predictors: 

\begin{description}
\item[Group 1:] Humidity: The mean relative humidity ($gm^{-3}$) over a 24-hour period.
\item[Group 2:] Wind speed: The maximum recorded wind speed ($mph$) within a 24-hour period.
\item[Group 3:] Precipitation: The total amount of rainfall ($mm$) within a 24-hour period.
\item[Group 4:] Lightning: The total number of lighting strikes within a 24-hour period.
\end{description}

The particular base transformation we consider is exponential smoothing, defined by
\begin{equation}\label{eqn:expsmooth}
x_{t,s} = \alpha x_{t,p} + (1-\alpha) x_{t,p} \quad \mbox{for}\ t=2,\dots,T
\end{equation}
where we set $x_{1,s}=x_{1,t}$.  In equation \eqref{eqn:expsmooth} the tuning parameter $\alpha$ is used to adjust how much the time series $x_{t,p}$ is
smoothed: a value of $\alpha$ close to 1 will produce a time series very close to the original, whilst a value
of $\alpha$ close to 0 will produce a time series that evolves much more slowly.  We apply the transformation to the predictors above for a range of values of $\alpha$, with the particular number and values chosen to sufficiently capture the non-linear effects for each predictor (guided by our industrial collaborator).  Note that due to the nature of the telecommunication events, all potential predictors should have a positive relationship to the response variables. The last group relates to indicator variables to adjust for calendar effects which are likely to influence the event data.  In particular we include three indicator variables, corresponding to the Christmas bank holiday (Christmas day and Boxing day); 27th December until New Year's Day; and any other bank holiday\footnote{Note that these variables are defined to adjust for those bank holidays which move from year to year.}.  

We present three methods for modelling the event data. The first method (denoted {\em Automated}) is our simultaneous predictor selection approach for multiple response variables, using our two-step procedure {\tt SPS2} to estimate a model for the regression residuals. The {\em Individual Automated} approach uses the two-step procedure of the first method, but is applied to each response variable separately. Consequently {\em Individual Automated} cannot take advantage of simultaneous predictor selection; we present this method to highlight the gains in a {\em simultaneous} predictor selection approach. Finally, {\em Current} is the procedure adopted by our industrial collaborator, included as a baseline comparison. This method removes the weekly seasonality and calendar effects from the response variables as part of a data pre-processing step, as these are not thought to be attributed to the effects of the predictors of interest (hence the bank holiday group of predictors is not considered for {\em Current}).  Data pre-processing choices can be subjective, as well as being time-consuming and therefore costly.  Furthermore, such pre-processing does not allow joint estimation of the external predictor, bank holiday effects and seasonality. Our two-step procedure for fitting a Reg-SARIMA model allows seasonality to be incorporated directly into the model specification which is iteratively updated as the predictor coefficients are refined. By modelling seasonality we can obtain more accurate estimates of prediction uncertainty and completely remove the need to pre-process the data by including calendar effects as indicator variables.

\begin{table*}[!h]
  \centering
  \resizebox{0.9\linewidth}{!}{
  \begin{tabular}{ll | ccc | ccc | ccc }
    & & \multicolumn{3}{c}{{\em Automated}}
    & \multicolumn{3}{c}{{\em Individual Automated}}
    & \multicolumn{3}{c}{{\em Current}} \\
    & & R1&R2&R3 & R1&R2&R3 & R1&R2&R3 \\
    \hline 
    \multirow{3}{*}{\shortstack{Group 1\\(humidity)}} & $\bs{\beta}_{1.1}$
      &-&-&- &-&-&- &-&-&- \\
    & $\bs{\beta}_{1.2}$
      &-&-&- &-&-&0.01 &-&-&0.01  \\
    &$\bs{\beta}_{1.3}$&0.01&0.01&0.01 &0.01&0.01&- &0.01&0.01&-  \\
    \hline
    \multirow{3}{*}{\shortstack{Group 2\\(wind)}} & $\bs{\beta}_{2.1}$
      &-&-&-&-&-&-&-&-&- \\
    & $\bs{\beta}_{2.2}$
      &0.02&0.02&0.01 &0.02&0.02&0.01 &0.03&0.02&0.03 \\
    & $\bs{\beta}_{2.3}$
      &-&-&- &-&-&- &-0.02&-0.02&-0.03  \\
    \hline
    \multirow{6}{*}{\shortstack{Group 3\\(rain)}} & $\bs{\beta}_{3.1}$
      &-&-&- &-&-&- &-0.03&-0.01&-0.02 \\
    & $\bs{\beta}_{3.2}$
      &-&-&- &-&-&- &0.21&1.12&0.13 \\
    & $\bs{\beta}_{3.3}$
      &0.06&0.05&0.05 &0.06&0.05& &-1.96&-4.55&-0.86 \\
    & $\bs{\beta}_{3.4}$
      &-&-&- &-&-&- &7&6.49&1.59 \\
    & $\bs{\beta}_{3.5}$
      &-&-&- &-&-&- &-9.87&-3.03&-0.77 \\
    & $\bs{\beta}_{3.6}$
      &-&-&- &-&-&0.09 &4.82&-0.00&-  \\
    \hline 
    \multirow{3}{*}{\shortstack{Group 4\\(lightning)}} & $\bs{\beta}_{4.1}$
      &-&-&- &-&-&- &-&-&0.01 \\
    & $\bs{\beta}_{4.2}$
      &-&-&-&-&-&-&-&-&- \\

    & $\bs{\beta}_{4.3}$
      &0.03&0.02&0.01 &0.03&0.02&0.01 &0.02&0.03&- \\
    \hline
    \multirow{3}{*}{\shortstack{Calendar\\effects}} & $\bs{\beta}_{5.1}$
      &-0.77&-0.78&-0.65 &-0.77&-0.78&-0.64 &-&-&- \\
    & $\bs{\beta}_{5.2}$
      &-0.73&-0.79&-0.68 &-0.73&-0.79&-0.68 &-&-&- \\
    & $\bs{\beta}_{5.3}$
      &-0.27&-0.27&0.24 &-0.27&-0.27&0.24 &-&-&- \\
  \end{tabular}
  }
  \caption{Regression coefficients for our proposed {\tt SPS2} method ({\em Automated}), our proposed {\tt SPS2} procedure applied to individual responses ({\em Individual Automated}) and the current implementation used by our industrial collaborator ({\em Current}). Each column represents the three different response variables in the dataset.  The rows represent the predictor variables.}
  \label{tab:BTcoefficients}
\end{table*}

The estimated regression coefficients for the three approaches are given in Table \ref{tab:BTcoefficients}. An immediate observation from Table \ref{tab:BTcoefficients} is that the models produced by the automated, two-step procedures ({\em Automated} and {\em Individual Automated} methods) are much sparser than those produced by the {\em Current} approach, not considering the calendar effects. Furthermore, all coefficients for the weather predictors produced from {\em Automated} and {\em Individual Automated} methods are positive, which, as outlined before, would be expected in this context for the telecommunications event data.  In contrast, the {\em Current} method includes highly correlated predictors, from the same group, and with opposing effects; for example, all six transformed variables of predictor 3 are included. Both large negative and large positive coefficients appear for the predictor variables from Group 3 for the {\em Current} method. This reflects the behavior of the least squares estimator discussed by \cite{HastieTibshiraniFriedmanESL} which motivated the use of the ridge penalty \citep{HoerlKennard1970}. Using simultaneous predictor selection and constraining the sign of the coefficients we are able to select the single best transformation of predictor 3. 

The mean squared errors for 14-day ahead predictions for the three methods are given in Table \ref{tab:msecomparisons}. The prediction accuracy is significantly reduced using the {\em Automated} and {\em Individual Automated} approaches that produce Reg-SARIMA models, rather than using a pre-processing step ({\em Current}). Recall that the Reg-SARIMA methods model the seasonality and calendar effects explicitly rather than remove it. They also describe the effects of other predictors. By selecting predictors simultaneously, the {\em Automated} approach provides more accurate forecasts of the response variables. We can see from Table \ref{tab:BTcoefficients} that different predictors from Groups 1 and 3 are chosen in comparison to Regions 1 and 2. 

\begin{table}[!h] \centering
  \begin{tabular}{r | ccc}
    & {\em Automated} & {\em Individual Automated} & {\em Current}\\
    \hline
    R1 & \textbf{0.204} & \textbf{0.204} & 0.280 \\
    R2 & \textbf{0.172} & 0.173 & 0.314 \\
    R3 & \textbf{0.173} & 0.182 & 0.212
  \end{tabular}
  \caption{Mean squared error for 14 day ahead predictions for each of the three response variables and the three methods described in the text.} \label{tab:msecomparisons}
\end{table}

\begin{figure*}[!h]
  \centering
  \begin{subfigure}{.32\linewidth}
    \includegraphics[width=\linewidth]
    {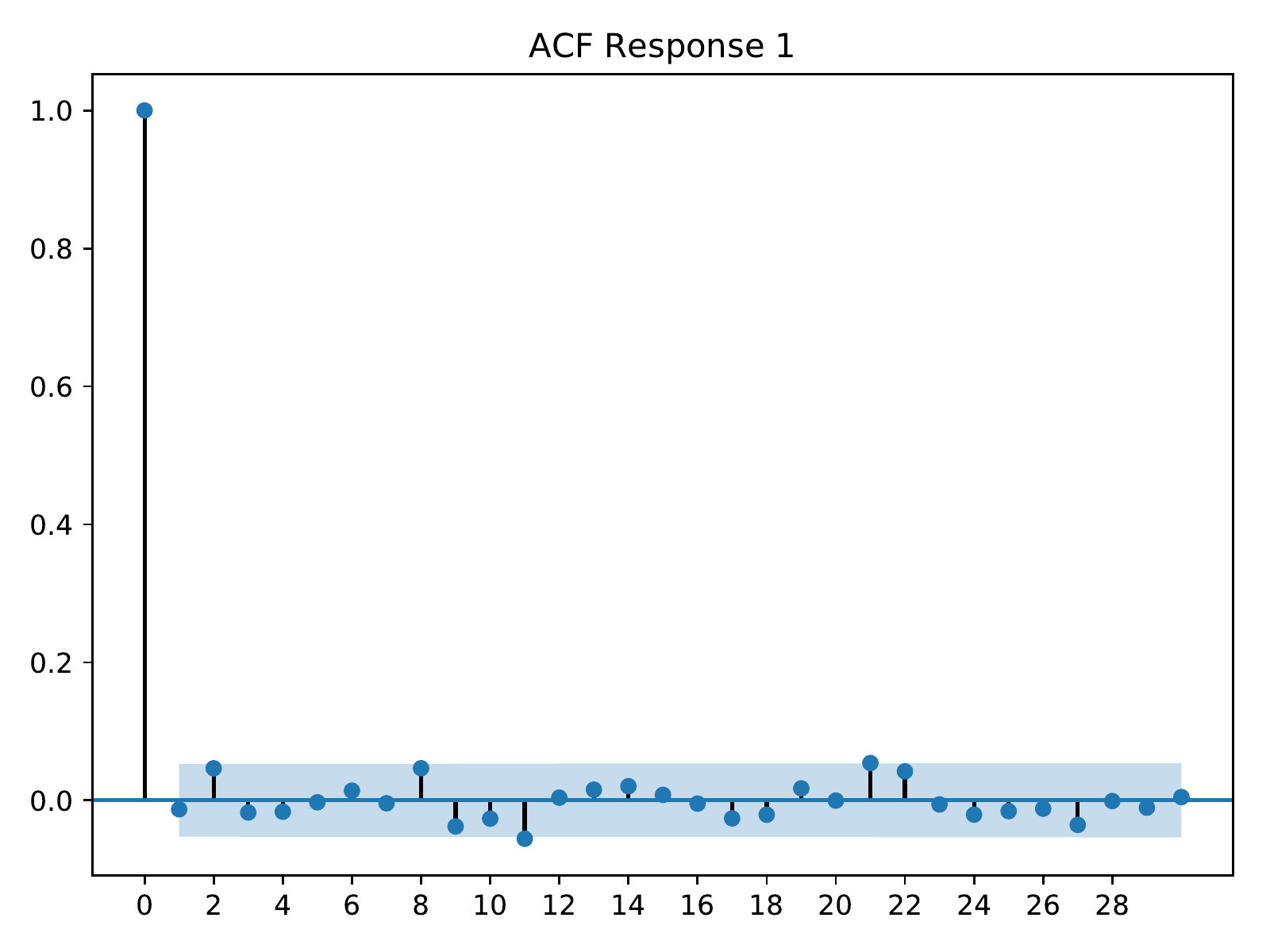}
  \end{subfigure}
  \begin{subfigure}{.32\linewidth}
    \includegraphics[width=\linewidth]
    {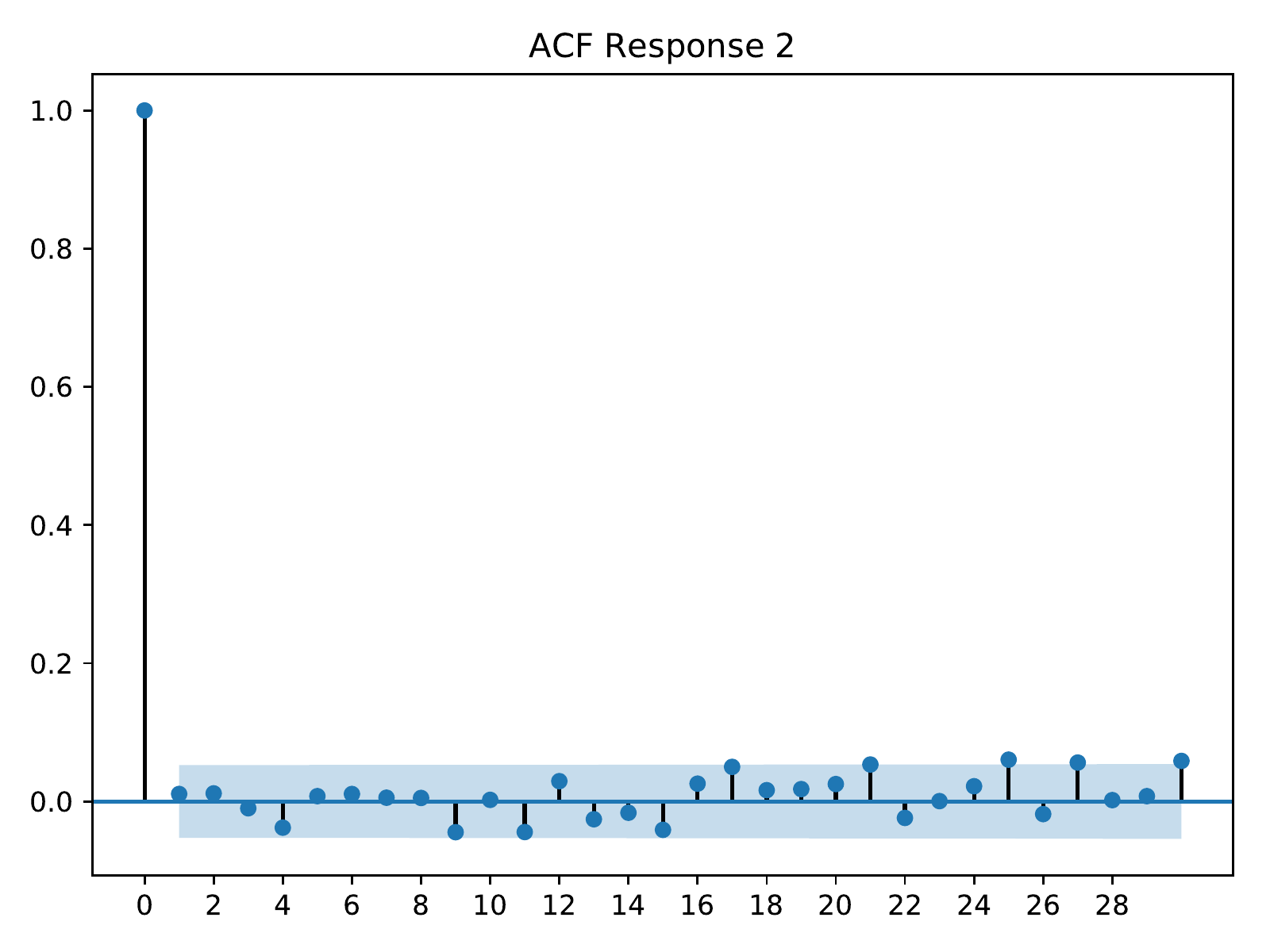}
  \end{subfigure}
  \begin{subfigure}{.32\linewidth}
    \includegraphics[width=\linewidth]
    {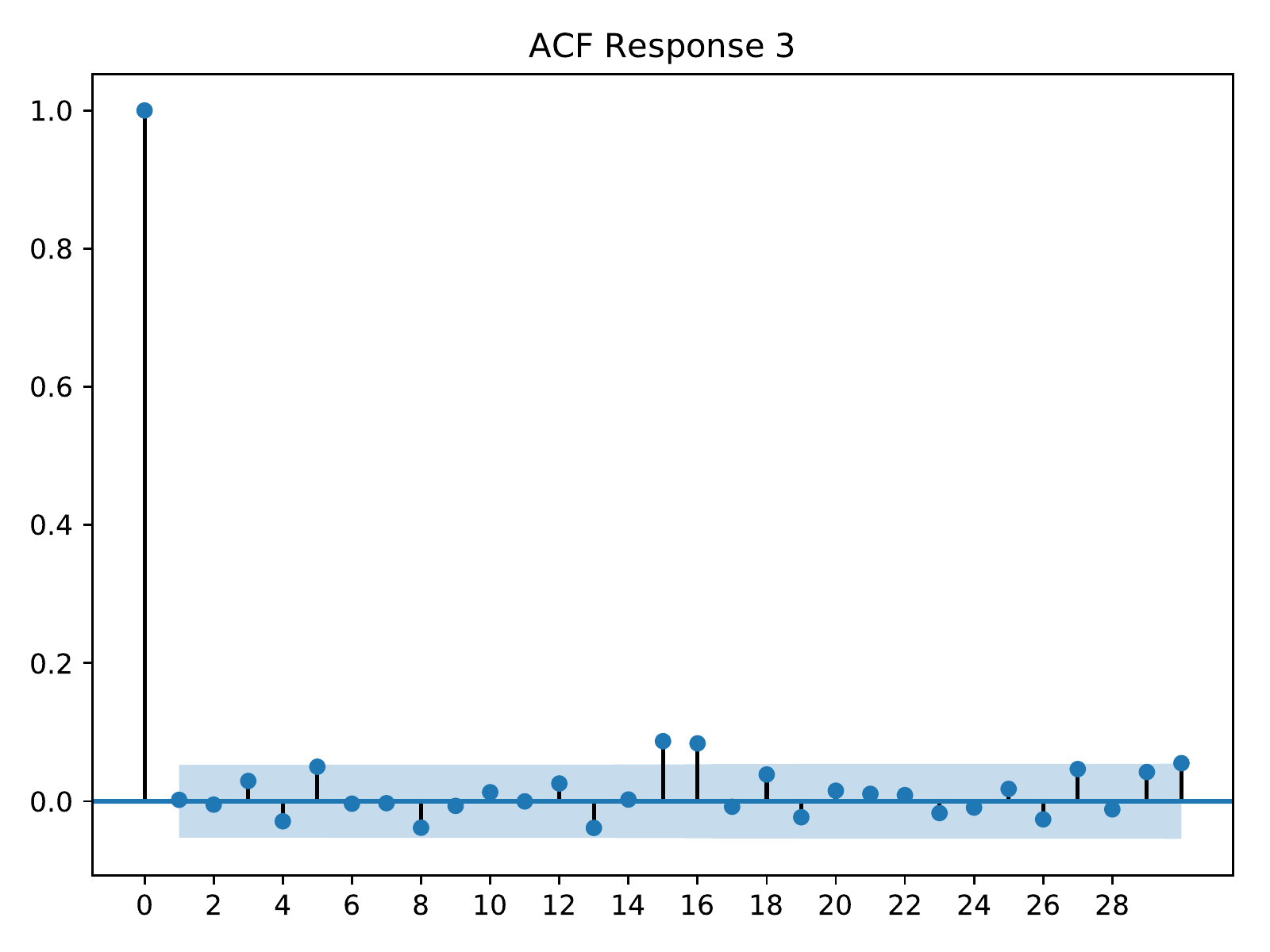}
  \end{subfigure}

  \begin{subfigure}{.32\linewidth}
    \includegraphics[width=\linewidth]
    {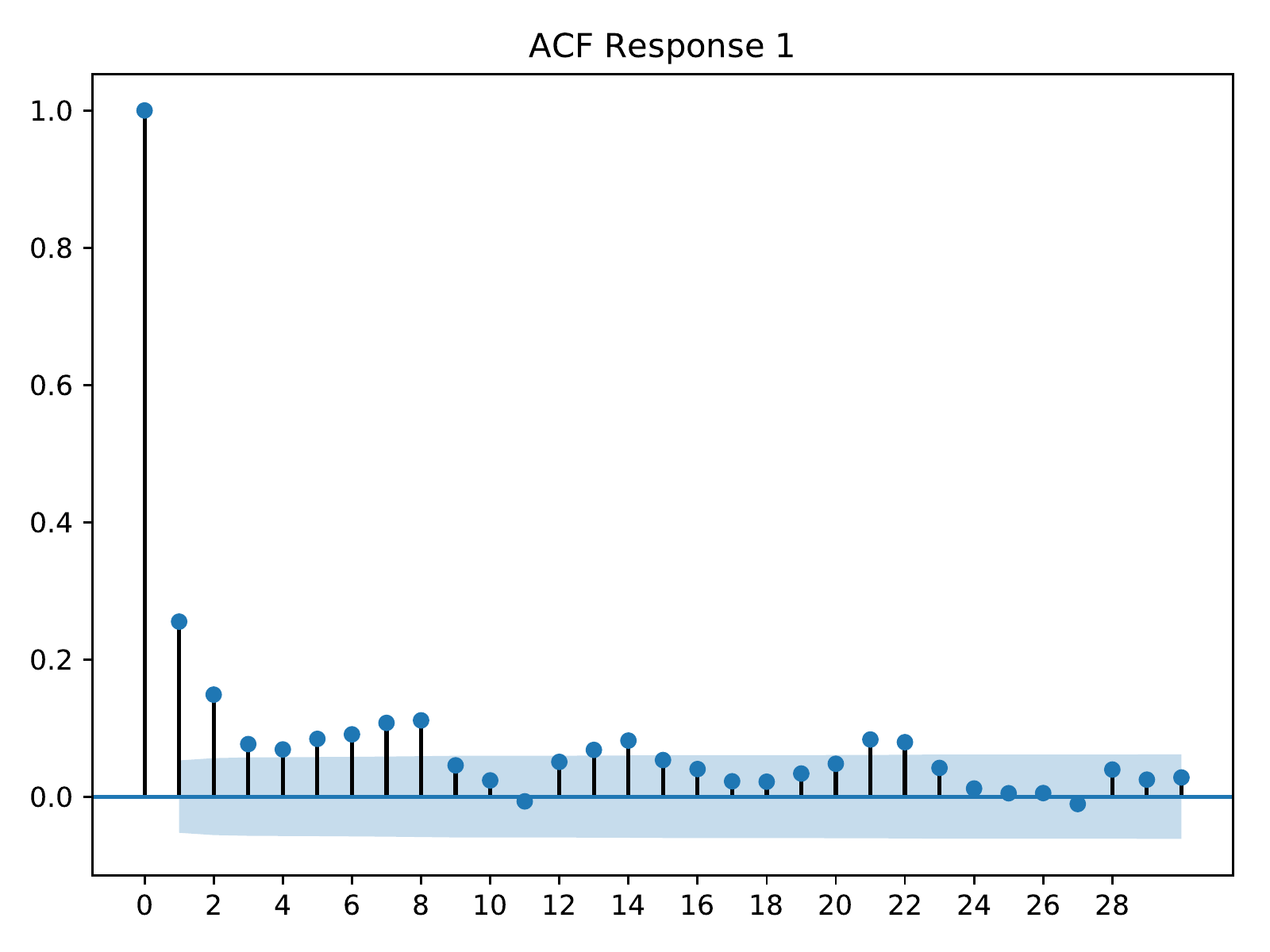}
    \caption{Response 1}
    \label{fig:acf-errors-current-approach-Response1}
  \end{subfigure}
  \begin{subfigure}{.32\linewidth}
    \includegraphics[width=\linewidth]
    {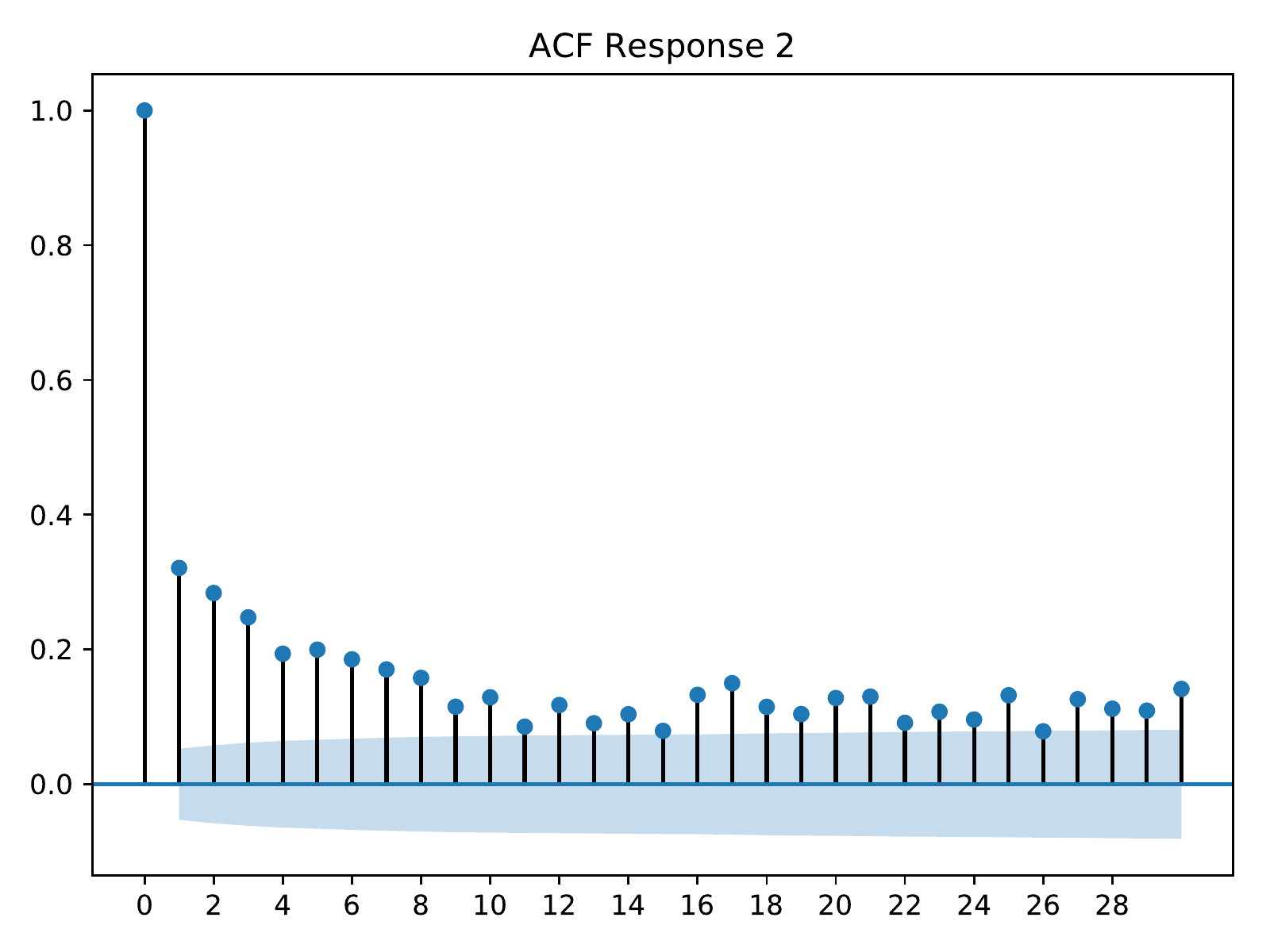}
    \caption{Response 2}
    \label{fig:acf-errors-current-approach-Response2}
  \end{subfigure}
  \begin{subfigure}{.32\linewidth}
    \includegraphics[width=\linewidth]{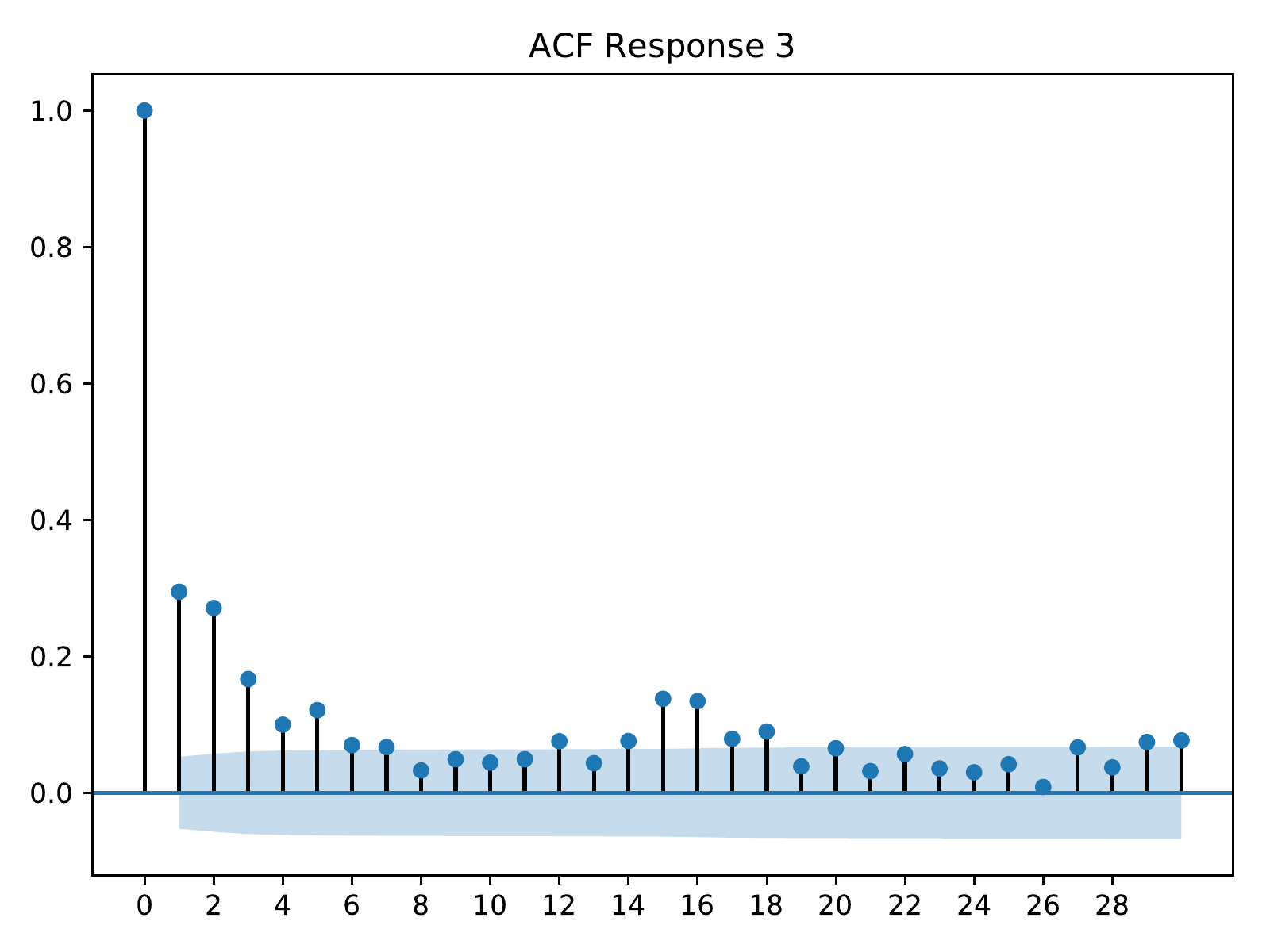}
    \caption{Response 3}
    \label{fig:acf-errors-current-approach-Response3} 
  \end{subfigure}
  \caption{The sample autocorrelation for the fitted model errors for each of the three response variables for the {\em Automated} method (top) and the {\em Current} method (bottom).}\label{fig:acf-errors}
\end{figure*}

To determine whether the SARIMA models produced by the {\em Automated} method have adequately captured the autocorrelation and seasonality within the data we can inspect the sample autocorrelation and sample partial autocorrelation functions of the model errors. The sample autocorrelation functions for the {\em Automated} and {\em Current} are shown in Figure \ref{fig:acf-errors}.

\begin{figure*}[!h]
  \centering
  \begin{subfigure}{.32\linewidth}
    \includegraphics[width=\linewidth]
    {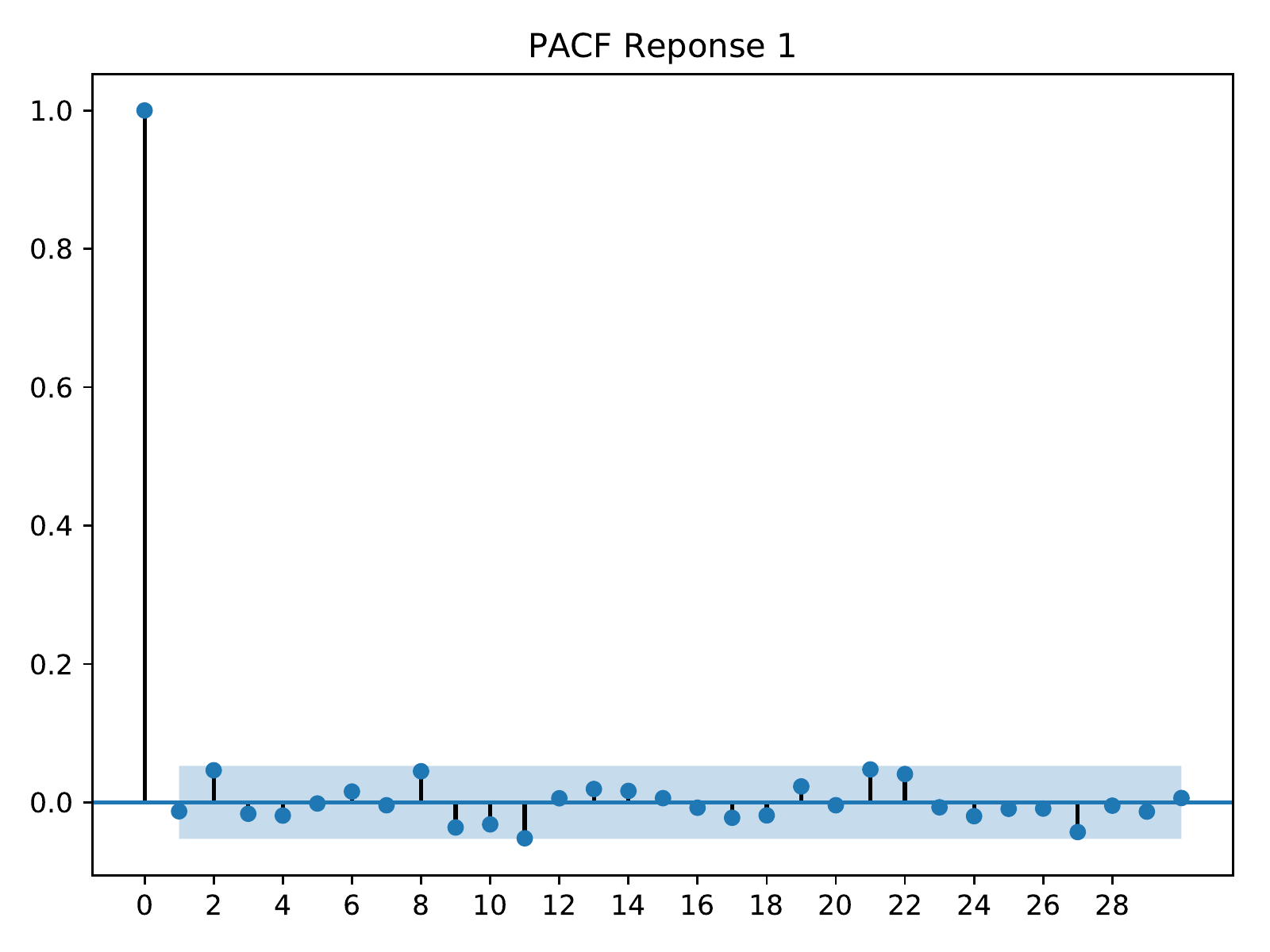}
  \end{subfigure}
  \begin{subfigure}{.32\linewidth}
    \includegraphics[width=\linewidth]
    {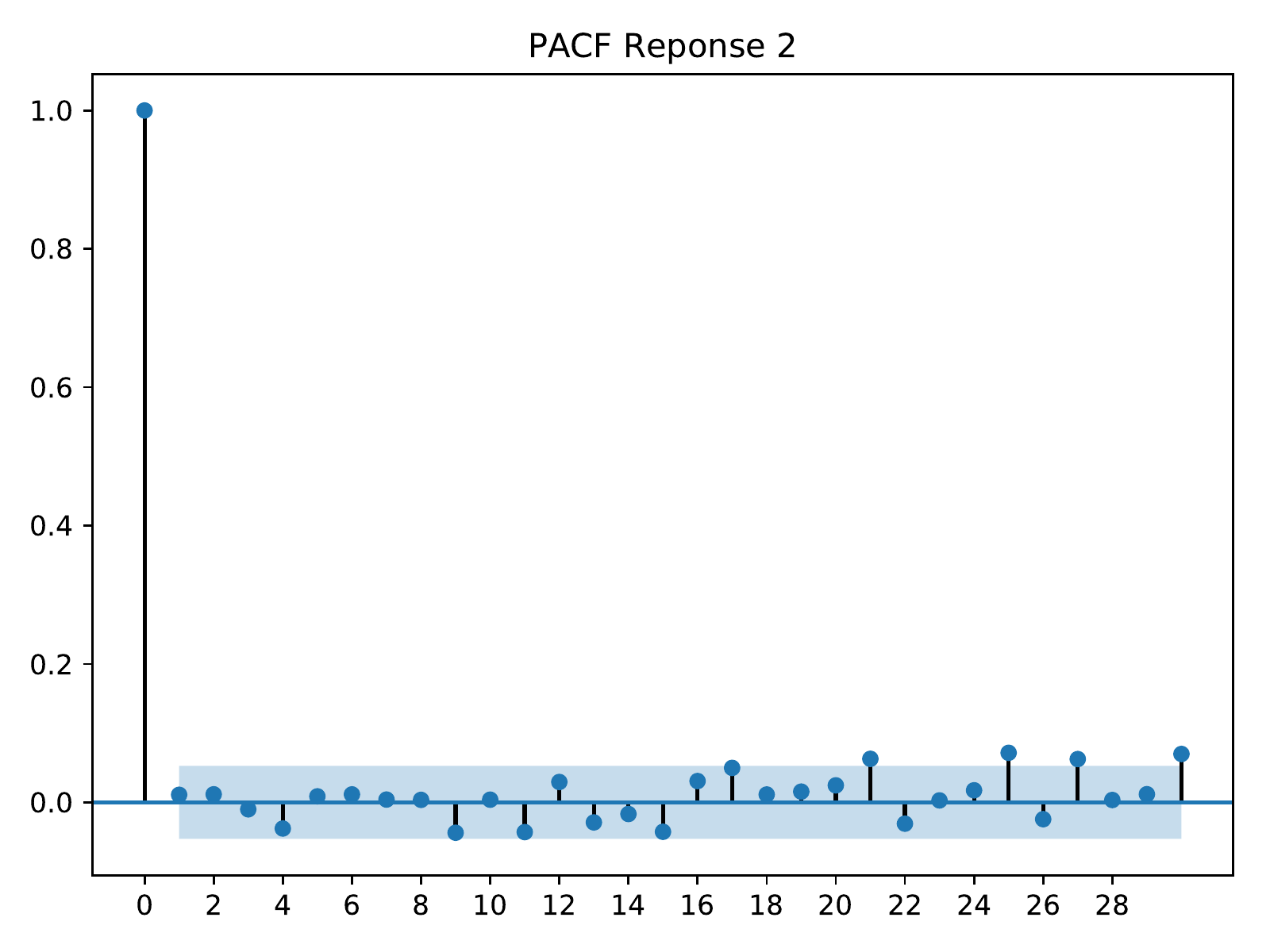}
  \end{subfigure}
  \begin{subfigure}{.32\linewidth}
    \includegraphics[width=\linewidth]
    {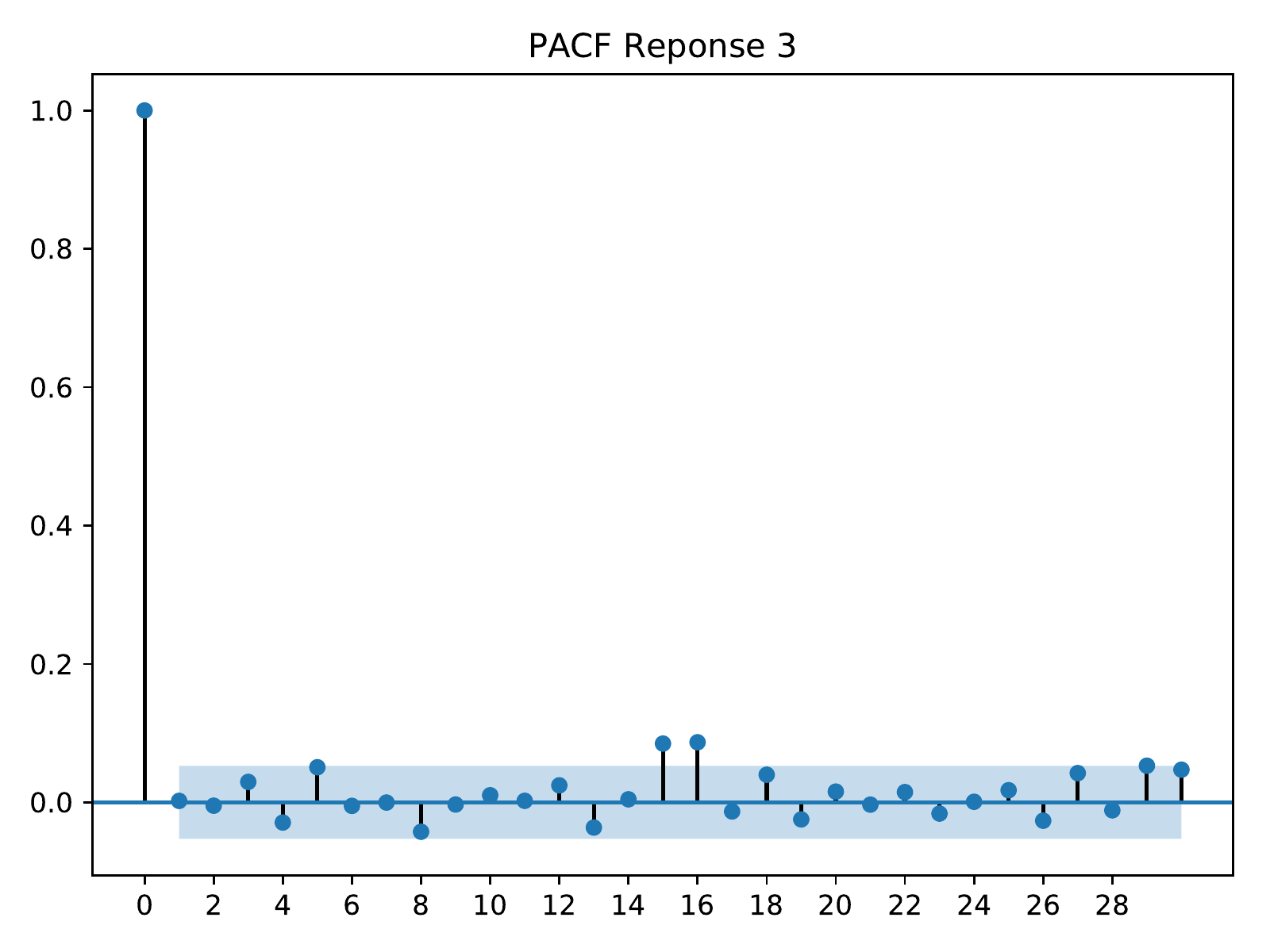}
  \end{subfigure}
  \begin{subfigure}{.32\linewidth}
    \includegraphics[width=\linewidth]
    {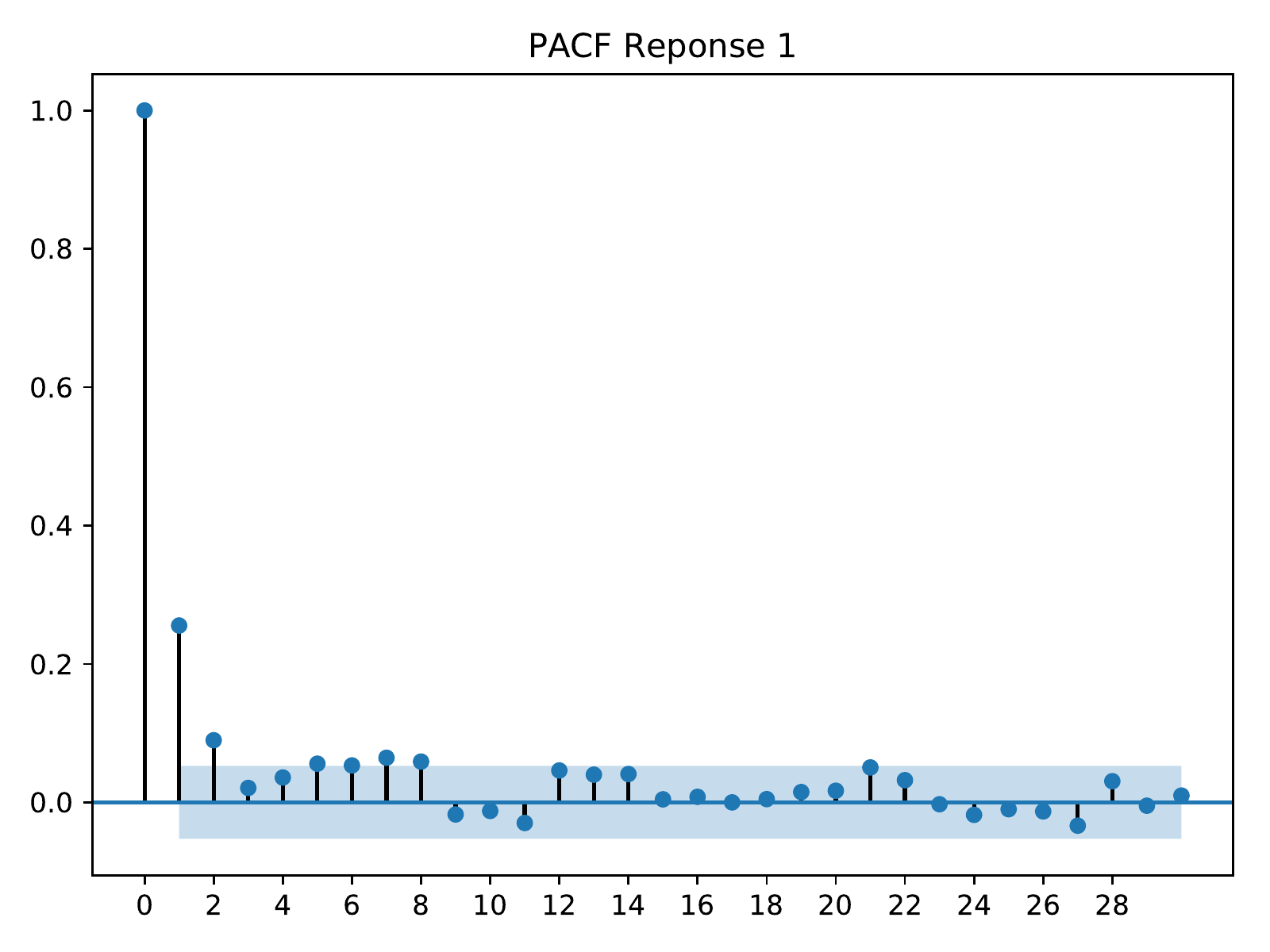}
    \caption{Response 1}
    \label{fig:pacf-errors-current-approach-Response1}
  \end{subfigure}
  \begin{subfigure}{.32\linewidth}
    \includegraphics[width=\linewidth]
    {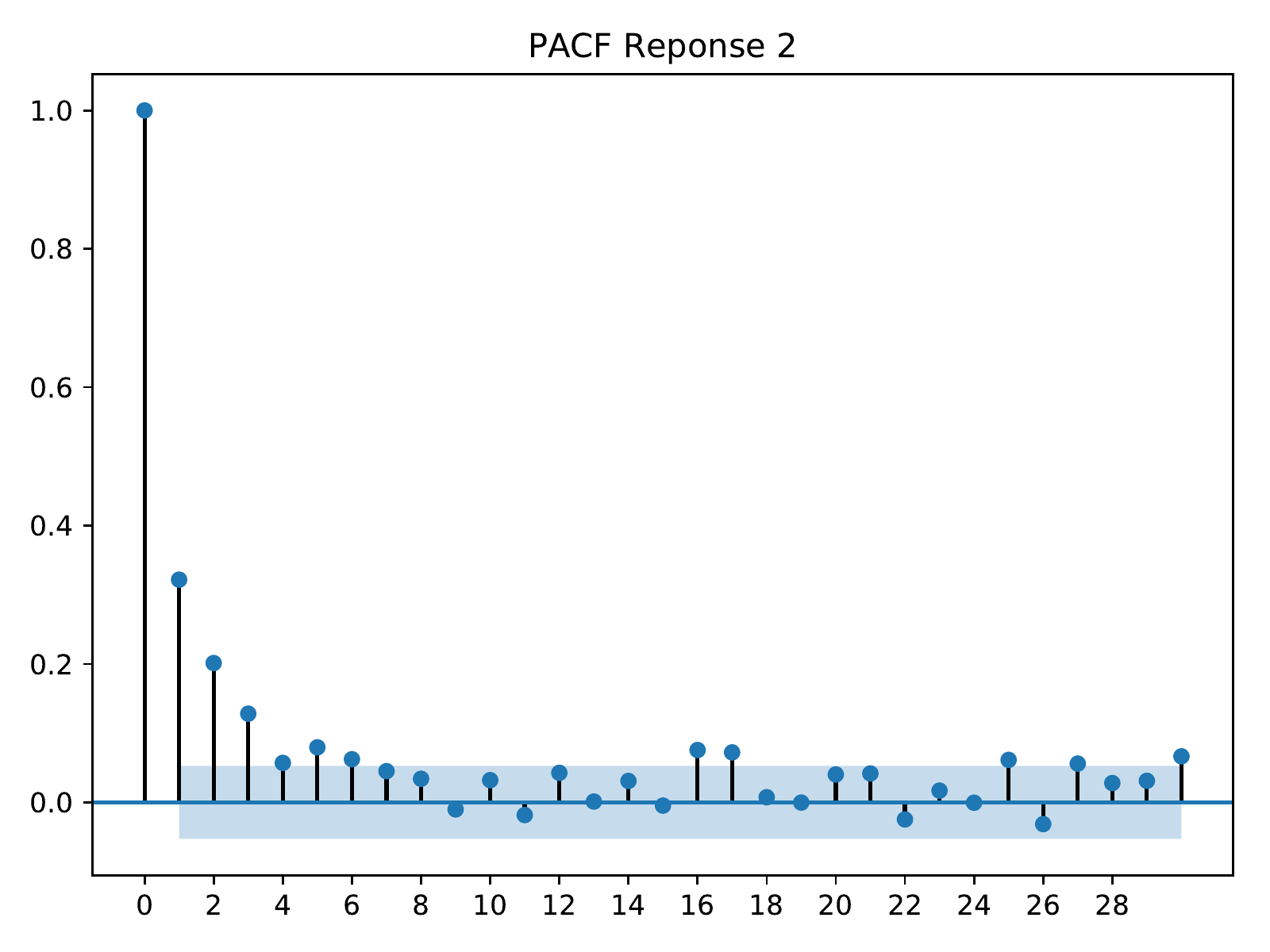}
    \caption{Response 2}
    \label{fig:pacf-errors-current-approach-Response2}
  \end{subfigure}
  \begin{subfigure}{.32\linewidth}
    \includegraphics[width=\linewidth]
    {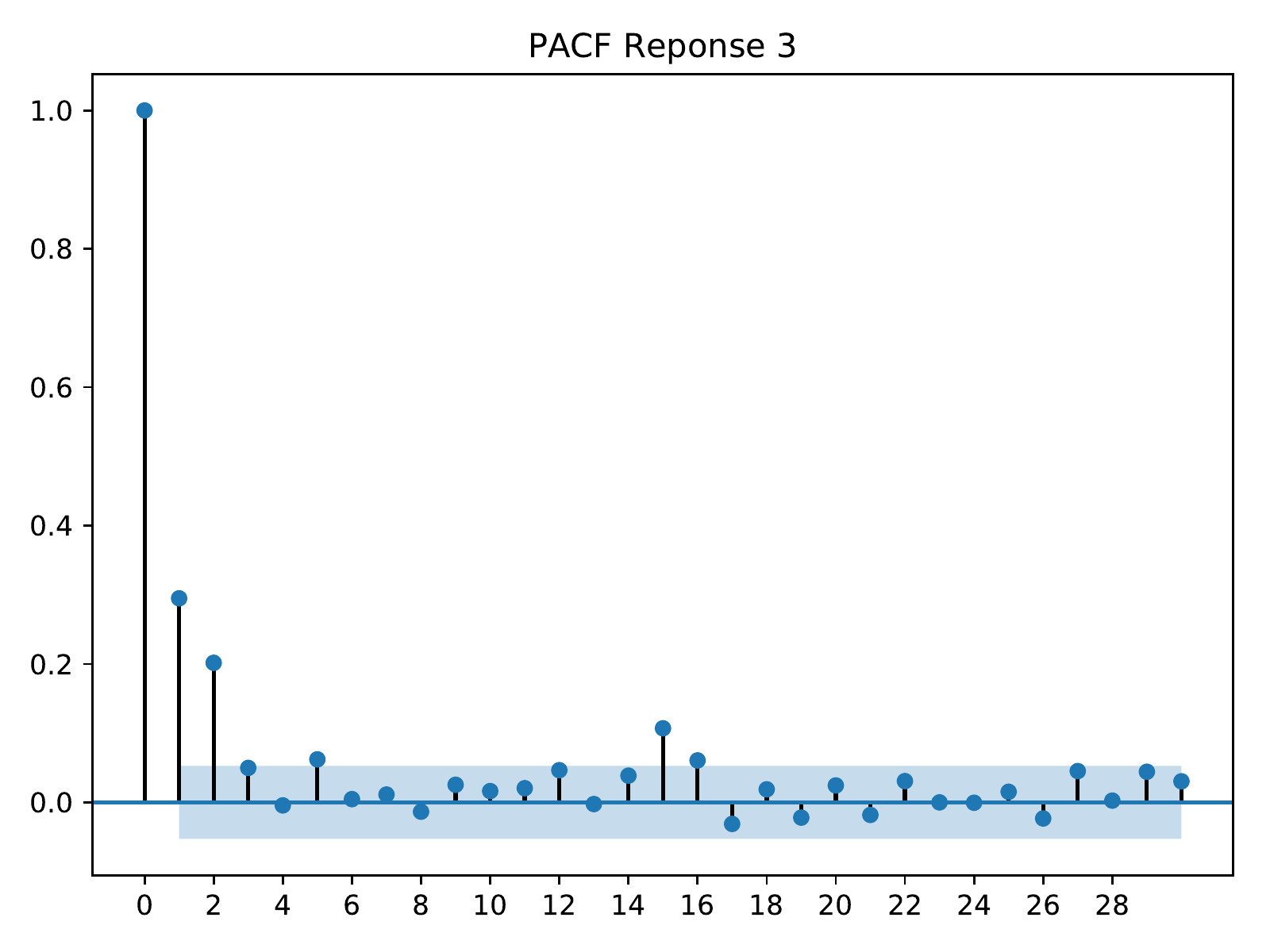}
    \caption{Response 3}
    \label{fig:pacf-errors-current-approach-Response3} 
  \end{subfigure}
  \caption{The sample partial autocorrelation for the fitted model errors for each of the three response variables for the {\em Automated} method (top) and the {\em Current} method (bottom).}
  \label{fig:pacf-errors}
\end{figure*}

The plots show that there is very little significant unmodelled autocorrelation left in the residuals for the {\em Automated} technique, demonstrating that modelling the regression residuals as a SARIMA process accounts for most of the temporal correlation (full model specifications for the {\em Automated} procedure can be found in Appendix \ref{sec:addit-data-appl}). In contrast, the {\em Current} method appears to violate the typical regression assumptions of independent regression residuals as there is significant remaining autocorrelation at many lags in the regression residuals for all three response variables. Similar conclusions can be drawn from the plots for the sample partial autocorrelation functions; these are shown in Figure \ref{fig:pacf-errors}. 

When serial correlation in the regression residuals is ignored the standard errors for each of the regression coefficients may be severely underestimated \citep{RawlingsPantulaSastryDickey}. This would raise suspicions about the significance of any predictor in the model. Further, prediction intervals are likely to be too narrow. Our observations mirror this tendency -- the standard errors of the regression coefficients for the three response variables produced from the {\em Automated} and {\em Current} methods are shown in Appendix \ref{sec:addit-data-appl}.  

\section{Concluding remarks}\label{sec:concl-furth-work}
Motivated by an industrial problem we have proposed a procedure to help automate the modelling process of telecommunications data. More specifically, we have developed a MIQO model to solve the simultaneous best subset problem to select predictors when jointly modelling multiple response variables. We have incorporated predictor selection within a two-step procedure, that iterates between selecting predictors for a regression model and modelling the serial correlation of the regression residuals. Automation is achieved by placing constraints in the MIQO formulation to ensure sensible models are produced, and by eliminating the need to pre-process the data through modelling calendar affects and seasonality. 

We have shown that predictor selection accuracy can be improved by simultaneously selecting predictors for multiple response variables. Selection accuracy and coefficient estimation can further be improved by shrinkage. The shrinkage we introduced is specifically designed for settings when joint estimation of models is considered -- in contrast to LASSO-like penalties that shrink coefficients towards zero, our shrinkage method forces coefficients between models to a common value.

Whilst not relevant for our dataset, an interesting avenue for future research would be to investigate the impact of modelling the regression residuals simultaneously. For example, in other settings modelling the regression residuals as a vector autoregression (VAR) could explain both temporal and cross correlations between the regression residuals between multiple responses. We anticipate that prediction error may be reduced further as well as give a consistent form for the regression residuals between responses.

\begin{acknowledgements}
Lowther gratefully acknowledges financial support from EPSRC and BT via the STOR-i Centre for Doctoral Training.

A Python implementation of the methods in this article will be publicly available in due course at\\\\ \texttt{https://github.com/aaronplowther/sps}.
\end{acknowledgements}

\appendix

 \section{Algorithmic details of the SVS-m procedure} \label{sec:addit-algor-deta}

In this section we outline the Simultaneous Variable Selection (SVS) problem and Convex Quadratic Program (CQP) used to solve it by \cite{TurlachVenablesWright2005}. The SVS problem was developed as an exploratory data analysis tool to determine suitable predictors for multiple response regression models. We modify the CQP to 
produce estimates for multiple regression models simultaneously. 

The problem proposed by \cite{TurlachVenablesWright2005} to acheive simultaneous predictor selection for multi-response models is
  \begin{equation}
    \label{eq:turlach-problem}
    \begin{aligned}
      \min_{\bs{\beta}} \sum_{m=1}^{M} \sum_{t=1}^{T}
      \left( y_{m,t} - \sum_{p=1}^{P} x_{p,t} \beta_{m,p} \right)^2 \quad \text{subject to}\\
      \sum_{p=1}^{P} \max \left( |\beta_{1,p}|, \ldots, |\beta_{M,p}| \right) \leq t.
    \end{aligned}
  \end{equation}

We propose a slight modification of this problem more suited for predictor selection in multiple separate linear regression models, as follows:
  \begin{equation}
    \label{eq:SMRMC-turlach-problem}
    \begin{aligned}
      \min_{\bs{\beta}} \sum_{m=1}^{M} \sum_{t=1}^{T}
      \left( y_{m,t} - \sum_{p=1}^{P} x_{m,p,t} \beta_{m,p} \right)^2 \quad \text{subject to}\\
      \sum_{p=1}^{P} \max \left( |\beta_{1,p}|, \ldots, |\beta_{M,p}| \right) \leq t.
    \end{aligned}
  \end{equation}

  Here, we allow a realisation of predictor $p$ for each response variable $Y_m$, for $m=1,\ldots,M$. The convex quadratic program formulated by 
\cite{TurlachVenablesWright2005} to solve (\ref{eq:turlach-problem}) is
  \begin{equation}\label{eq:turlach-formulation}
    \begin{aligned}
      \min_{\bs{\beta}} \sum_{m=1}^{M} \sum_{t=1}^{T}
      \left( y_{m,t} - \sum_{p=1}^{P} x_{p,t} \beta_{m,p} \right)^2 \quad \text{subject to}\\
      \bs{u}_{M} \otimes \bs{z} - \bs{\beta} \geq 0 \\
      \bs{u}_{M} \otimes \bs{z} + \bs{\beta} \geq 0 \\
      t - \bs{u}_P \bs{z} \geq 0,
    \end{aligned}
  \end{equation}
  
where $\bs{u}_M \in \mathbb{R}^M$ with each entry equal to 1, and $\bs{z} \in \mathbb{R}^P$ give auxiliary variables. We modify formulation 
(\ref{eq:turlach-formulation}) to give
  \begin{equation}
    \label{eq:smrmcturlach-formulation}
    \begin{aligned}
      \min_{\bs{\beta}} \sum_{m=1}^{M} \sum_{t=1}^{T}
      \left( y_{m,t} - \sum_{p=1}^{P} x_{m,p,t} \beta_{m,p} \right)^2 \quad \text{subject to}\\
      \bs{u}_{M} \otimes \bs{z} - \bs{\beta} \geq 0 \\
      \bs{u}_{M} \otimes \bs{z} + \bs{\beta} \geq 0 \\
      t - \bs{u}_P \bs{z} \geq 0.
    \end{aligned}
  \end{equation}

The final step is to determine the maximum value $t$. We set $t_{\text{max}} = \sum_{m=1}^{M} \sum_{p=1}^{P} |\hat{\beta}_{m,p}|$ where $\hat{\beta}_{m,p}$ minimises\newline $\sum_{m=1}^{M} \sum_{t=1}^{T}
  \left( y_{m,t} - \sum_{p=1}^{P} x_{m,p,t} \beta_{m,p} \right)^2$. All coefficients which are solutions to formulation (\ref{eq:smrmcturlach-formulation}) are non zero. 
We apply the same heuristic proposed by \cite{TurlachVenablesWright2005} to determine those that should be zero. Let
  \[
    \mathcal{I} = \{p : || \bs{\beta}_p ||_{\infty} > t \times e^{-4}, \ \text{for} \ p = 1,\ldots,P\}.
  \]
  All coefficients $\beta_{m,p}$ for $p \notin \mathcal{I}$ are set to zero. 

\section{Additional simulation results}\label{app:simstudy}

This appendix provides additional simulation results for our proposed algorithm described in the main article.  We also summarise some computational aspects of the algorithm.


\subsection{Simulated performance of SBS for increasing residual variance}

As the variance of the regression residuals increases the variation in the response variables is increasingly attributed to randomness rather than changes in the 
predictors. This makes it much harder to recover the true predictors. To investigate this, we simulate $N=1000$ synthetic datasets and compare the predictor selection 
accuracy and mean squared error of the regression coefficients between the best subset method, applied independently to the $M$ regression models, and our 
{\em simultaneous} best subset implementation. The level of sparsity, $k=3$, is set to the true model sparsity.  

\begin{figure}[!h]
  \centering 
    \includegraphics[width=0.95\linewidth]
    {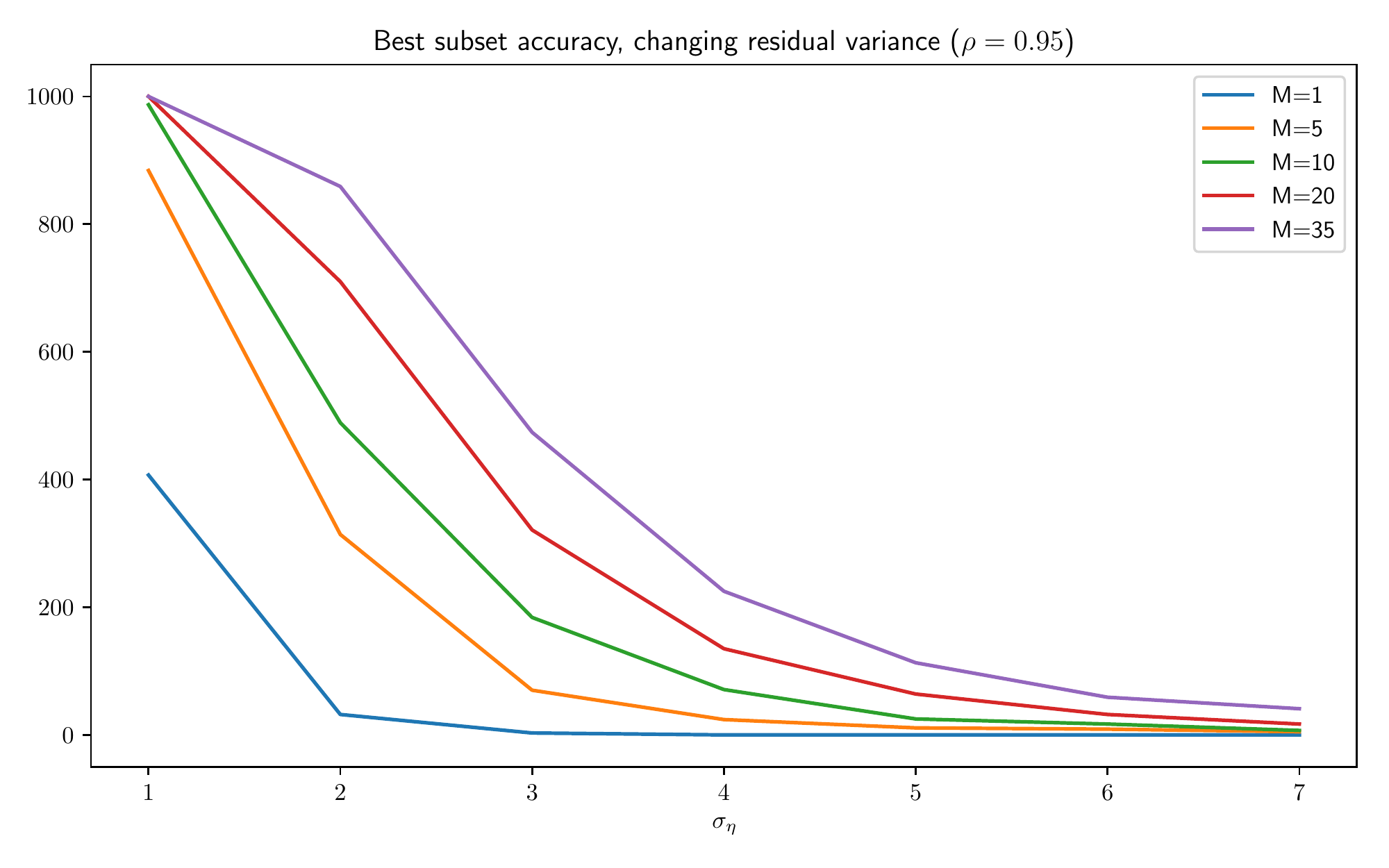}
  \caption{Selection accuracy of SBS as the residual variance $\sigma^2_\eta$ increases for different numbers of jointly-modelled response variables, $M$. \label{fig:numcorrect-increasing-M-sigma}}
\end{figure}

Figure \ref{fig:numcorrect-increasing-M-sigma} shows the selection accuracy for increasing $\sigma^2_{\eta}$ for different values of jointly-modelled responses, $M$. With $\rho=0.95$ we observe the deterioration in selection accuracy as $\sigma_{\eta}^2$ increases and how this is improved by the SBS approach.  In particular, Figure \ref{fig:numcorrect-increasing-M-sigma} shows that the best subset method is unable to recover the true predictors for $\sigma_{\eta}^2 \geq 3$. Improvements in predictor selection can be achieved by increasing $M$, and there appears to be improved consistency in the SBS approach over the best subset approach as $M$ increases. 

\begin{figure}[!h]
  \centering 
    \includegraphics[width=0.95\linewidth]
    {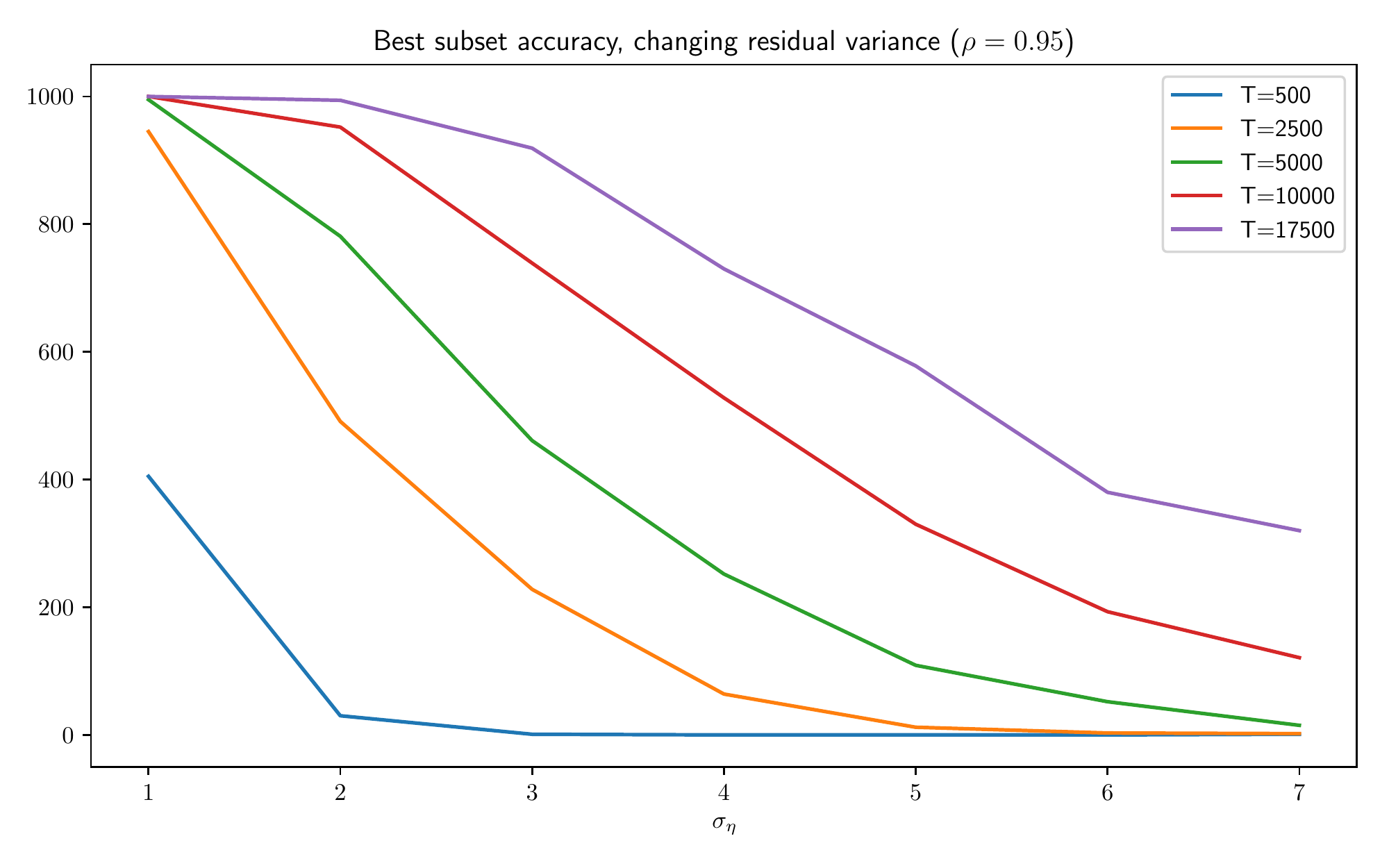}
  \caption{Selection accuracy of SBS as the residual variance $\sigma^2_\eta$ increases for increasing dataset length, $T$. \label{fig:numcorrect-increasing-T-univariate}}
\end{figure}

We also compare the selection accuracy for SBS applied with $M$ response variables each with $T$ observations in Figure \ref{fig:numcorrect-increasing-T-univariate} to the best subset method ($M=1$) with $MT$ observations. Our gain in selection accuracy is not quite as high as previously observed. However, in many applications of interest (for example our motivating telecommunications setting) it is not always possible to increase the number of observations. 

\subsection{Computational aspects} \label{sec:comp-aspects}

We now illustrate the total runtime of the solver in worst case scenarios. In this simulation study, all data is simulated from the multi-response regression
model where the predictors have the specification
\[
  X \sim \text{MVN}_P(\bs{0}, \bs{\Sigma}) \quad \text{where} \
  \Sigma \in \mathbb{R}^{P \times P}:= \Sigma_{i,j} = 0.25^{|i-j|}.
\]
The number of response variables, $M$ and the number of predictors, $P$ will be made clear where relevant. The regression coefficients are given by
\[
  \beta_{m,p} =
  \begin{cases}
    1, \quad \text{if} \ p = 1,3,5, \\
    0, \quad \text{otherwise},
  \end{cases} \ \text{for} \ m=1,\ldots,M.
\]
In each simulation we use $T=500$ observations.

\begin{figure}[!h]
  \centering
  \includegraphics[width=0.95\linewidth]{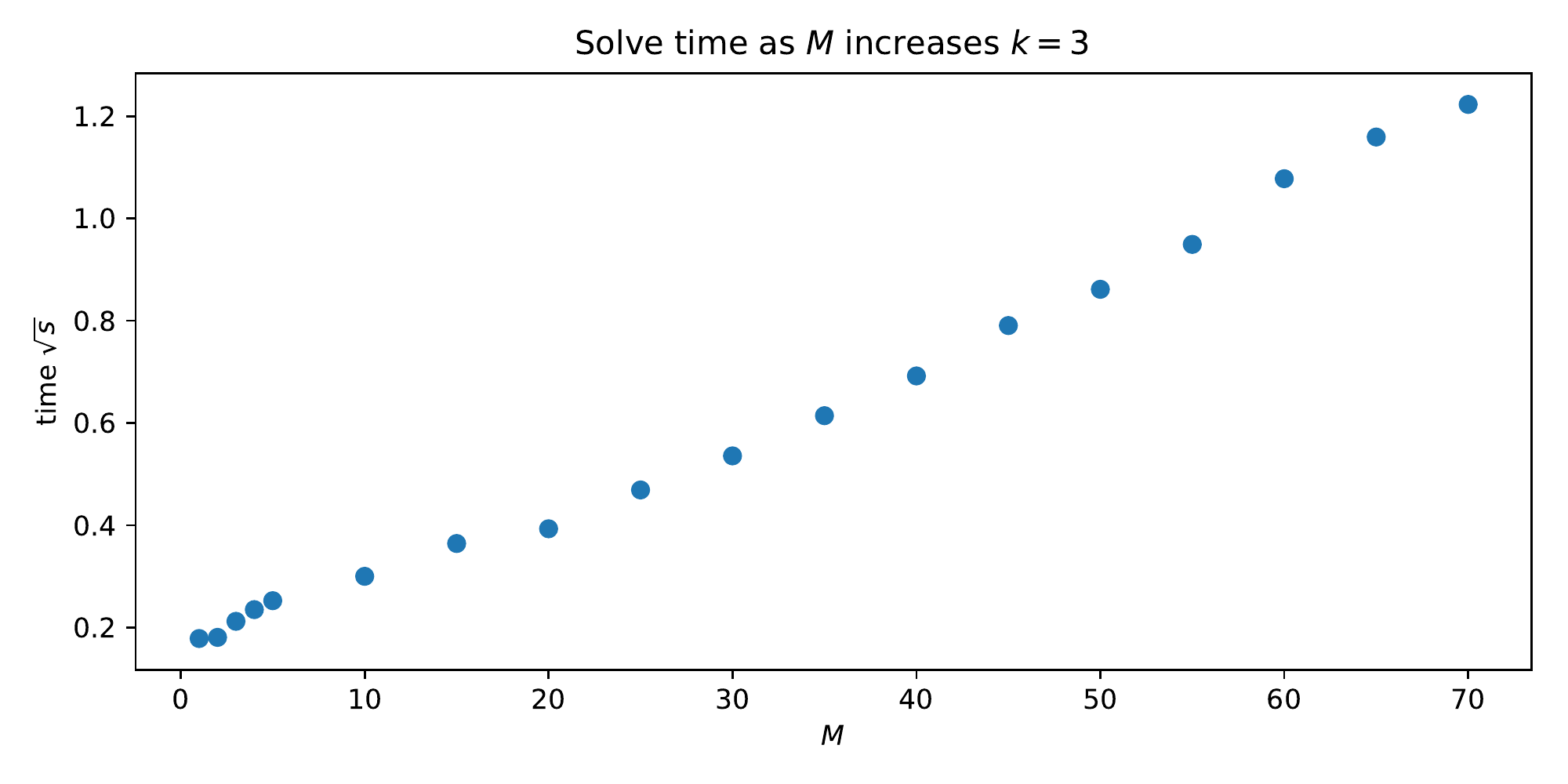}
  \caption{Solve time (square root scale) for SMRMC as $M$, the number of response variables increases. Here $k=3$ and $P=30$.}
  \label{fig:solvetime-increasing-M}
\end{figure}

Firstly, we investigate the impact of increasing $M$ on our formulation for the SBS problem with $P=30$ predictors. The number of continuous variables in the SBS problem increases linearly with $M$ and causes the total solve time to increase quadratically. This can be seen in Figure \ref{fig:solvetime-increasing-M}. 

\begin{figure}[!h]
  \centering
  \includegraphics[width=0.95\linewidth]{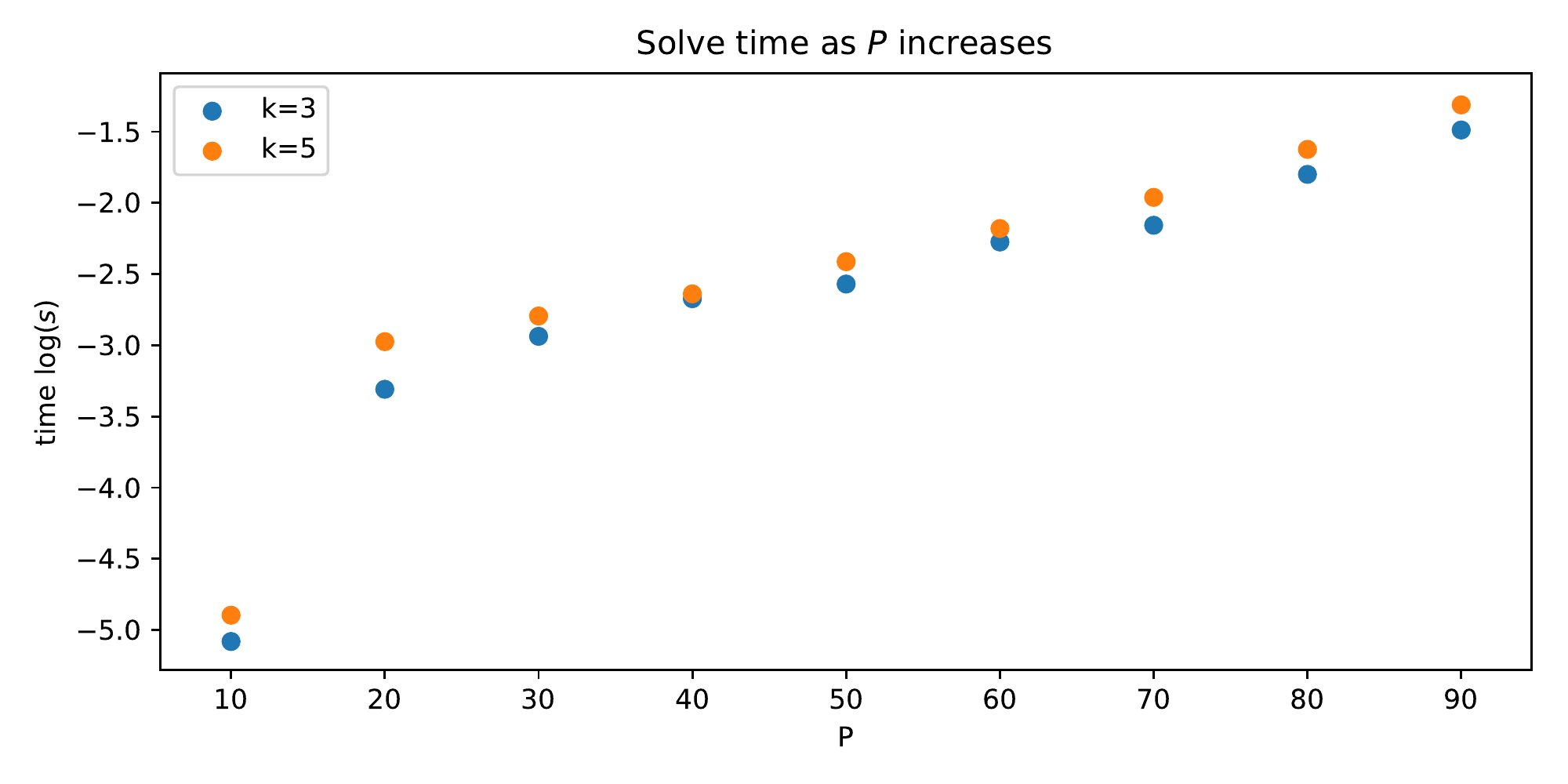}
  \caption{Solve time (log scale) for SMRMC as $P$, the number of predictors increases. Here $M=5$ and results are shown for $k=3$ and $5$.}
  \label{fig:solvetime-increasing-P}
\end{figure}

There is a near linear trend for the solve time on the square root scale. As the number of predictors, $P$ increases, the number of continuous and binary variables increase linearly. For a fixed $k$, the possible number of predictor combinations increases exponentially. The total solve times are shown in Figure \ref{fig:solvetime-increasing-P}. For a fixed level of sparsity the total solve time increases exponentially with $P$.

\section{Additional model details from the telecommunications data study}\label{sec:addit-data-appl}

In this section we provide the full specification of Reg-SARIMA models produced by the {\em Automated} method for the three response variables.  More specifically, the model coefficients, standard errors, 
and lower and upper quartiles (LQ and UQ respectively) of the model coefficients for the {\em Automated} approach are shown in Tables \ref{tab:summary_3regions_Broadlands} -- \ref{tab:summary_3regions_CI}. For the {\em Current} method, the results are shown in Tables \ref{tab:SummaryKjeld1} to ~\ref{tab:SummaryKjeld3}.

\begin{table}[!ht]
  \centering
  \begin{tabular}{l | cccc}
    Parameter & Estimate & Standard error & LQ & UQ \\
    \hline
    $\beta_{1,1.3}$ & 0.131 & 0.028 & 0.076 & 0.185 \\ \hline
    $\beta_{1,2.2}$ & 0.119 & 0.022 & 0.077 & 0.162 \\ \hline
    $\beta_{1,3.3}$ & 0.154 & 0.020 & 0.114 & 0.194 \\ \hline
    $\beta_{1,4.3}$ & 0.189 & 0.012 & 0.165 & 0.212 \\ \hline
    $\beta_{1,5.1}$ & 0.200 & 0.016 & 0.168 & 0.231 \\
    $\beta_{1,5.2}$ & 0.149 & 0.023 & 0.104 & 0.194 \\
    $\beta_{1,5.3}$ & 0.081 & 0.029 & 0.025 & 0.138 \\ \hline
    $\phi_{1,1}$ & 0.914 & 0.136 & 0.648 & 1.180 \\
    $\phi_{1,2}$  & -0.082 & 0.049 & -0.178 & 0.014 \\
    $\theta_{1,1}$ & -0.709 & 0.132 & -0.967 & -0.450 \\
    $\Phi_{1,1}$ & 0.057 & 0.032 & -0.006 & 0.119 \\
    $\Theta_{1,1}$ & -0.979 & 0.009 & -0.996 & -0.962 \\ \hline
    $\sigma_{\eta,1}$ & 0.293 & 0.007 & 0.279 & 0.307 \\
  \end{tabular}
  \caption{Model summary for Response 1 with Reg-SARIMA model order (2,0,1)(1,1,1,7).}
  \label{tab:summary_3regions_Broadlands}
\end{table}

\begin{table}[!ht]
  \centering
  \begin{tabular}{l | cccc}
    Parameter & Estimate & Standard error & LQ & UQ \\
    \hline
    $\beta_{2,1.3}$ & 0.151 & 0.036 & 0.081 & 0.221 \\ \hline
    $\beta_{2,2.2}$ & 0.094 & 0.018 & 0.058 & 0.130 \\ \hline
    $\beta_{2,3.3}$ & 0.128 & 0.024 & 0.080 & 0.176 \\ \hline
    $\beta_{2,4.3}$ & 0.094 & 0.022 & 0.051 & 0.136 \\ \hline
    $\beta_{2,5.1}$ & 0.189 & 0.017 & 0.155 & 0.223 \\
    $\beta_{2,5.2}$ & 0.157 & 0.023 & 0.113 & 0.202 \\
    $\beta_{2,5.3}$ & 0.081 & 0.027 & 0.029 & 0.133 \\ \hline
    $\phi_{2,1}$   & 0.797 & 0.045 & 0.708 & 0.885 \\
    $\theta_{2,1}$ & -0.597 & 0.053 & -0.701 & -0.492 \\
    $\Phi_{2,1}$   & 0.043 & 0.024 & -0.005 & 0.091 \\
    $\Theta_{2,1}$ & -0.964 & 0.009 & -0.981 & -0.946 \\ \hline
    $\sigma_{\eta,2}$ & 0.298 & 0.006 & 0.286 & 0.310 \\
  \end{tabular}
  \caption{Model summary for Response 2 with Reg-SARIMA model order (1,0,1)(1,1,1,7).}
  \label{tab:summary_3regions_Cambridge}
\end{table}

\begin{table}[!ht]
  \centering
  \begin{tabular}{l | cccc}
    Parameter & Estimate & Standard error & LQ & UQ \\
    \hline
    $\beta_{3,1.3}$ & 0.125 & 0.040 & 0.047 &  0.203 \\ \hline
    $\beta_{3,2.2}$ & 0.083 & 0.026 & 0.031 &  0.133 \\ \hline
    $\beta_{3,3.3}$ & 0.133 & 0.034 & 0.067 &  0.199 \\ \hline
    $\beta_{3,4.3}$ & 0.090 & 0.027 & 0.037 &  0.144 \\ \hline
    $\beta_{3,5.1}$ & 0.166 & 0.014 & 0.138 &  0.193 \\
    $\beta_{3,5.2}$ & 0.137 & 0.026 & 0.086 &  0.189 \\
    $\beta_{3,5.3}$ & 0.077 & 0.032 & 0.014 &  0.140 \\ \hline
    $\phi_{3,1}$   & 0.221 & 0.012 & 0.197 &  0.245 \\
    $\theta_{3,1}$ & 0.180 & 0.013 & 0.155 &  0.205 \\
    $\Phi_{3,1}$   & 0.022 & 0.009 & 0.004 &  0.040 \\
    $\Theta_{3,1}$ & -0.972 & 0.010 & -0.991 & -0.952 \\ \hline
    $\sigma_{\eta,3}$ & 0.397 & 0.006 & 0.386 & 0.408 \\
  \end{tabular}
  \caption{Model summary for Response 3 with Reg-SARIMA model order (2,0,0)(1,1,1,7).}
  \label{tab:summary_3regions_CI}
\end{table}

\begin{table}[!ht]
\centering
\begin{tabular}{l | cccc}
Parameter & Estimate & Standard Error & LQ & UQ \\
\hline
$\beta_{0,1.3}$ & 0.009 & 0.002 & 0.006 & 0.012 \\ \hline
$\beta_{0,2.2}$ & 0.027 & 0.006 & 0.016 & 0.039 \\
$\beta_{0,2.3}$ & -0.025 & 0.008 & -0.040 & -0.001 \\ \hline
$\beta_{0,3.1}$ & -0.029 & 0.005 & -0.039 & -0.019 \\
$\beta_{0,3.2}$ & 0.209 & 0.055 & 0.101 & 0.316 \\
$\beta_{0,3.3}$ & -1.962 & 0.801 & -3.532 & -0.391 \\
$\beta_{0,3.4}$ & 7.0 & 2.833 & 1.447 & 12.553 \\
$\beta_{0,3.5}$ & -9.872 & 3.774 & -17.269 & -2.475 \\
$\beta_{0,3.6}$ & 4.824 & 1.715 & 1.4617 & 8.186 \\ \hline
$\beta_{0,4.3}$ & 0.017 & 0.003 & 0.0111 & 0.024 \\
\end{tabular}
\caption{Model summary for Response 1 from the standard approach used by our industrial collaborator.}
\label{tab:SummaryKjeld1}
\end{table}

\begin{table}[!ht]
\centering
\begin{tabular}{l | cccc}
Parameter & Estimate & Standard Error & LQ & UQ \\
\hline
$\beta_{1,1.3}$ & 0.013 & 0.007 & -0.001 & 0.026 \\ \hline
$\beta_{1,2.2}$ & 0.025 & 0.002 & 0.022 & 0.028 \\
$\beta_{1,2.3}$ & -0.017 & 0.005 & -0.026 & -0.007 \\ \hline
$\beta_{1,3.1}$ & -0.011 & 0.006 & -0.023 & 0.001 \\
$\beta_{1,3.2}$ & 1.120 & 0.003 & 1.115 & 1.126 \\
$\beta_{1,3.3}$ & -4.550 & 0.313 & -5.163 & -3.936 \\
$\beta_{1,3.4}$ & 6.493 & 1.464 & 3.623 & 9.363 \\
$\beta_{1,3.5}$ & -3.030 & 2.238 & -7.416 & 1.357 \\
$\beta_{1,3.6}$ & -0.003 & 1.104 & -2.166 & 2.161 \\ \hline
$\beta_{1,4.3}$ & 0.033 & 0.003 & 0.028 & 0.038 \\
\end{tabular}
\caption{Model summary for Response 2 from the standard approach used by our industrial collaborator.}
\label{tab:SummaryKjeld2}
\end{table}

\begin{table}[!ht]
\centering
\begin{tabular}{l | cccc}
Parameter & Estimate & Standard Error & LQ & UQ \\
\hline
$\beta_{2,1.2}$ & 0.007 & 0.001 & 0.004 & 0.009 \\ \hline
$\beta_{2,2.2}$ & 0.028 & 0.007 & 0.015 & 0.041 \\
$\beta_{2,2.3}$ & -0.028 & 0.009 & -0.045 & -0.011 \\ \hline
$\beta_{2,3.1}$ & -0.020 & 0.005 & -0.029 & -0.011 \\
$\beta_{2,3.2}$ & 0.133 & 0.034 & 0.055 & 0.211 \\
$\beta_{2,3.3}$ & -0.865 & 0.322 & -1.496 & -0.233 \\
$\beta_{2,3.4}$ & 1.589 & 0.656 & 0.303 & 2.876 \\
$\beta_{2,3.5}$ & -0.769 & 0.384 & -1.522 & -0.016 \\ \hline
$\beta_{2,4.3}$ & 0.015 & 0.003 & 0.009 & 0.020 \\
\end{tabular}
\caption{Model summary for Response 3 from the standard approach used by our industrial collaborator.}
\label{tab:SummaryKjeld3}
\end{table}

\bibliographystyle{spbasic}

\bibliography{References}

\end{document}